\documentclass[a4paper]{revtex4-2}

\usepackage[utf8]{inputenc}
\usepackage[english]{babel}
\usepackage{graphics}
\usepackage{graphicx}
\usepackage{hyperref}
\usepackage{amsmath}
\usepackage{amssymb}
\usepackage{siunitx}
\usepackage{physics}

\usepackage{csquotes}
\usepackage{bm}
\usepackage{comment}
\usepackage{sidecap}

\begin{document}

\begin{center}
\textit{\textbf{Accepted author manuscript:} This manuscript has been accepted for publication in \\Philosophical Transactions of the Royal Society A under the DOI} \texttt{10.1098/rsta.2025.0087}.
\end{center}
\vspace{1cm}

\title{Global stability of the Atlantic overturning circulation:\\ Edge state, long transients and boundary crisis under CO$_2$ forcing}

\author{Reyk Börner}
\email{r.borner@uu.nl}
\affiliation{Department of Mathematics and Statistics, University of Reading, Reading, UK}
\affiliation{Institute for Marine and Atmospheric Research Utrecht, Utrecht University, Utrecht, The Netherlands}

\author{Oliver Mehling}
\affiliation{Department of Environment, Land and Infrastructure Engineering, Politecnico di Torino, Turin, Italy}
\affiliation{Institute for Marine and Atmospheric Research Utrecht, Utrecht University, Utrecht, The Netherlands}

\author{Jost von Hardenberg}
\affiliation{Department of Environment, Land and Infrastructure Engineering, Politecnico di Torino, Turin, Italy}

\author{Valerio Lucarini}
\affiliation{School of Computing and Mathematical Sciences, University of Leicester, Leicester, UK}

\date{August 29, 2025}

\begin{abstract}
    The Atlantic Meridional Overturning Circulation (AMOC), a crucial ocean current system, could transition to a weak state. Despite severe associated climate impacts, assessing the AMOC's response under global warming and its proximity to possible critical thresholds remains difficult. To understand future Earth system stability, a global dynamical view is needed beyond the local stability analysis underlying classical early-warning methods. Using an intermediate-complexity climate model, we explore the stability landscape of the AMOC for different atmospheric CO$_2$ concentrations. We explicitly compute the edge state (or Melancholia state), a chaotic saddle on the basin boundary separating the strong and weak AMOC attractors found in the model. While being unstable, the edge state can govern the transient climate for centuries, supporting centennial AMOC oscillations driven by atmosphere-ice-ocean interactions in the North Atlantic. At increased CO$_2$ levels projected for the near future, we reveal a boundary crisis where the current AMOC attractor disappears by colliding with the edge state. Under crisis overshoot, long chaotic transients due to ``ghost states'' lead to diverging ensemble trajectories under time-varying forcing. Rooted in dynamical systems theory, our results offer an explanation of large ensemble variance and apparent ``stochastic bifurcations'' observed in earth system models under intermediate forcing scenarios. 
\end{abstract}

\maketitle
\onecolumngrid

\section{Introduction}
Earth's climate is a metastable complex system \cite{rossi_dynamical_2025}: on various scales, the variability of paleoclimate records is characterised by relatively abrupt transitions between distinct long-lived climatic regimes \cite{westerhold_astronomically_2020,dansgaard_evidence_1993,alley_abrupt_2003}. From a dynamical systems perspective, we may interpret the observed metastability by regarding the Earth system as a forced multistable system featuring a hierarchy of competing attracting states \cite{feudel_complex_2008,margazoglou_dynamical_2021}. The stability landscape of the underlying time-frozen system is thereby described by a global quasipotential based on Graham's field theory \cite{graham_nonequilibrium_1986,graham_nonequilibrium_1991}, with local minima of the landscape corresponding to attractors.

The quasipotential landscape of the Earth system has been explored in the context of the Cenozoic Era \cite{rousseau_punctuated_2023} and for our planet's multistable extent of glaciation (\enquote{Snowball Earth}) \cite{lucarini_edge_2017,lucarini_global_2020,margazoglou_dynamical_2021}. Here, we close in on a subscale feature of the present-day climate thought to be multistable: the Atlantic Meridional Overturning Circulation (AMOC), 
a widely studied system of large-scale ocean currents \cite{lynch-stieglitz_atlantic_2017, rahmstorf_ocean_2002, henry_north_2016}. 
The AMOC plays a vital role in climate by transporting heat northwards, supporting northern Europe's relatively mild climate \cite{weijer_stability_2019}. A suspected driver of past climate metastability \cite{lynch-stieglitz_atlantic_2017}, the AMOC is one of the proposed climate tipping elements \cite{armstrong_mckay_exceeding_2022, ashwin_tipping_2012}. Given the ongoing anthropogenic climate change \cite{masson-delmotte_climate_2021}, there is growing concern that one or more tipping elements could cross a \textit{tipping point} and transition to a qualitatively different state, with severe consequences for humanity and nature \cite{lenton_global_2023, wunderling_climate_2024}. The possibility of tipping events complicates climate prediction, contributing large uncertainty to risk management and adaptation strategies. 

How the AMOC will respond to global warming is an urgent open question. Climate models forced with greehnouse gas emissions scenarios until the year 2100 consistently project a decline of the AMOC, though its magnitude is model-dependent \cite{weijer_cmip6_2020,masson-delmotte_climate_2021}. While some studies infer a recent weakening from reconstructions \cite{caesar_observed_2018}, direct measurements are short and noisy \cite{mccarthy_sustainable_2020}.

An AMOC shutdown would have severe global impacts, including a relative cooling of the northern hemisphere, reduction in precipitation and increased storminess in Europe, shifts of rainfall patterns globally, and regional accelerations in sea level rise \cite{liu_overlooked_2017,bellomo_future_2021}. Even without a full shutdown, a partial collapse of the circulation could result from the shutdown of deep convection zones in the North Atlantic Subpolar Gyre (SPG) \cite{levermann_bistability_2007,sgubin_abrupt_2017}. Such a transition could cause qualitatively similar impacts within decades \cite{swingedouw_risk_2021}.

A hierarchy of climate models -- from box models \cite{stommel_thermohaline_1961} to intermediate-complexity \cite{rahmstorf_thermohaline_2005, hawkins_bistability_2011} and comprehensive earth system models \cite{jackson_hysteresis_2018, van_westen_asymmetry_2023} -- indicates that the AMOC can be multistable \cite{broecker_does_1985, manabe_two_1988, manabe_are_1999}. In a certain regime of atmospheric heat and freshwater forcing, a vigorous flow state resembling today's circulation (ON state) coexists with a much weaker or collapsed overturning state (OFF state). While there could be additional competing states \cite{lohmann_multistability_2024}, this bistability underlies the classical view of the AMOC as a tipping element. The bistability stems from the positive salt-advection feedback, describing the interdependence between the AMOC flow strength and northward salt transport \cite{kuhlbrodt_driving_2007,weijer_stability_2019}, which could be triggered by surface buoyancy changes in the North Atlantic.

To address the risk of an AMOC transition, research has aimed at detecting early warning signs (EWS), determining critical forcing levels, and estimating tipping times. A series of recent studies \cite{boers_observation-based_2021,michel_early_2022,ditlevsen_warning_2023} has applied statistical EWS to time series of observed AMOC reconstructions, suggesting that the AMOC is approaching a tipping point that could be reached this century \cite{ditlevsen_warning_2023}. These methods rely on the concept of critical slowing-down as the system approaches a bifurcation. An indicator based on the salt import into the Atlantic supported these findings when applied to reanalysis data \cite{van_westen_physics-based_2024}.
However, these methods have serious limitations that make the robust prediction of a transition difficult in practice, if not prohibitive \cite{ben-yami_uncertainties_2023}. A crucial underlying assumption is that the system remains close to an equilibrium state, which may not hold given the current rate of anthropogenic forcing. Instead, the scenario of nonautonomous or rate-dependent tipping \cite{ashwin_tipping_2012, wieczorek_rate-induced_2023} is more appropriate, for which an EWS theory is missing and indicators based on critical slowing-down fail \cite{huang_deep_2024}.

The trajectory of the Earth system constitutes one realization of a chaotic complex system \cite{lorenz_deterministic_1963}. This implies limits to predictability intrinsically linked with chaos \cite{lorenz_climatic_1975,mehling_limits_2024}. Near critical thresholds of metastable systems, the sensitive dependence on the initial condition (predictability of the first kind \cite{lorenz_climatic_1975}) can strongly inhibit the predictability of the asymptotic state (second kind) \cite{knutti_limited_2002,lohmann_predictability_2024}. Particularly, an ensemble of trajectories may partially tip under identical time-dependent forcing, simply due to internal variability: some ensemble members transition, while others do not. Ensemble splitting has been found in climate models of intermediate and high complexity \cite{kaszas_snowball_2019, romanou_stochastic_2023, lohmann_predictability_2024, gu_wide_2024}. In the NASA GISS-E2-1-G Earth system model (hereafter GISS model) of the Coupled Model Intercomparison Project Phase 6 (CMIP6), an ensemble sampled from internal variability showed divergent AMOC behaviour described as a \enquote{stochastic bifurcation} \cite{romanou_stochastic_2023}.
By construction, critical slowing-down indicators cannot discern between such trajectories that tip and those that do not \cite{huang_deep_2024}. To understand this behaviour and predict tipping in chaotic nonautonomous systems, a \textit{global} stability view beyond stable equilibria is needed.

While studies often emphasise the binary question of tipping or not tipping, the transient behaviour can be equally important \cite{lai_transient_2011, grebogi_crises_1983}. Long transients and metastable dynamics are often governed by unstable states (non-attracting invariant sets), which are underexplored in climate models \cite{feudel_rate-induced_2023}. A particular class of unstable states are \textit{edge states}, also called \textit{Melancholia states} \cite{lucarini_edge_2017,lohmann_melancholia_2024}: saddles embedded in the basin boundaries that partition the state space between the competing stable states. Edge states may be defined as attractors of the dynamics restricted to the basin boundary -- pictorially, \enquote{mountain passes} between valleys of the global quasipotential landscape \cite{borner_climate_2025}. Thus, edge states often act as gateways of critical transitions (with caveats \cite{borner_saddle_2024}). Numerically, edge states can be found using an edge tracking algorithm originally proposed in the context of turbulent flows \cite{battelino_multiple_1988,skufca_edge_2006,schneider_laminar-turbulent_2008} and recently applied to climate models \cite{lucarini_edge_2017,mehling_limits_2024,lohmann_melancholia_2024,borner_climate_2025}.

In complex systems, the basin boundaries may exhibit a highly complicated geometry \cite{mehling_limits_2024,bodai_rough_2020,lucarini_edge_2017}. Rather than marking a sharp threshold, the boundary can be a gray zone in state space of fractal dimension where the system is virtually unpredictable. In parameter space, this leads to a tipping window, as opposed to a sharp tipping point, initiated when an attracting state collides with the boundary in a so-called boundary crisis \cite{grebogi_crises_1983,ashwin_contrasting_2025}.
In the tipping window, the system may undergo long transients, such that a transition might occur thousands of years after a loss of stability \cite{drijfhout_shutdown_2025}. These aspects highlight that determining a precise tipping threshold or timing may not be possible in finite time, requiring a more holistic assessment of stability.

\begin{SCfigure}
    \includegraphics[width=0.5\columnwidth]{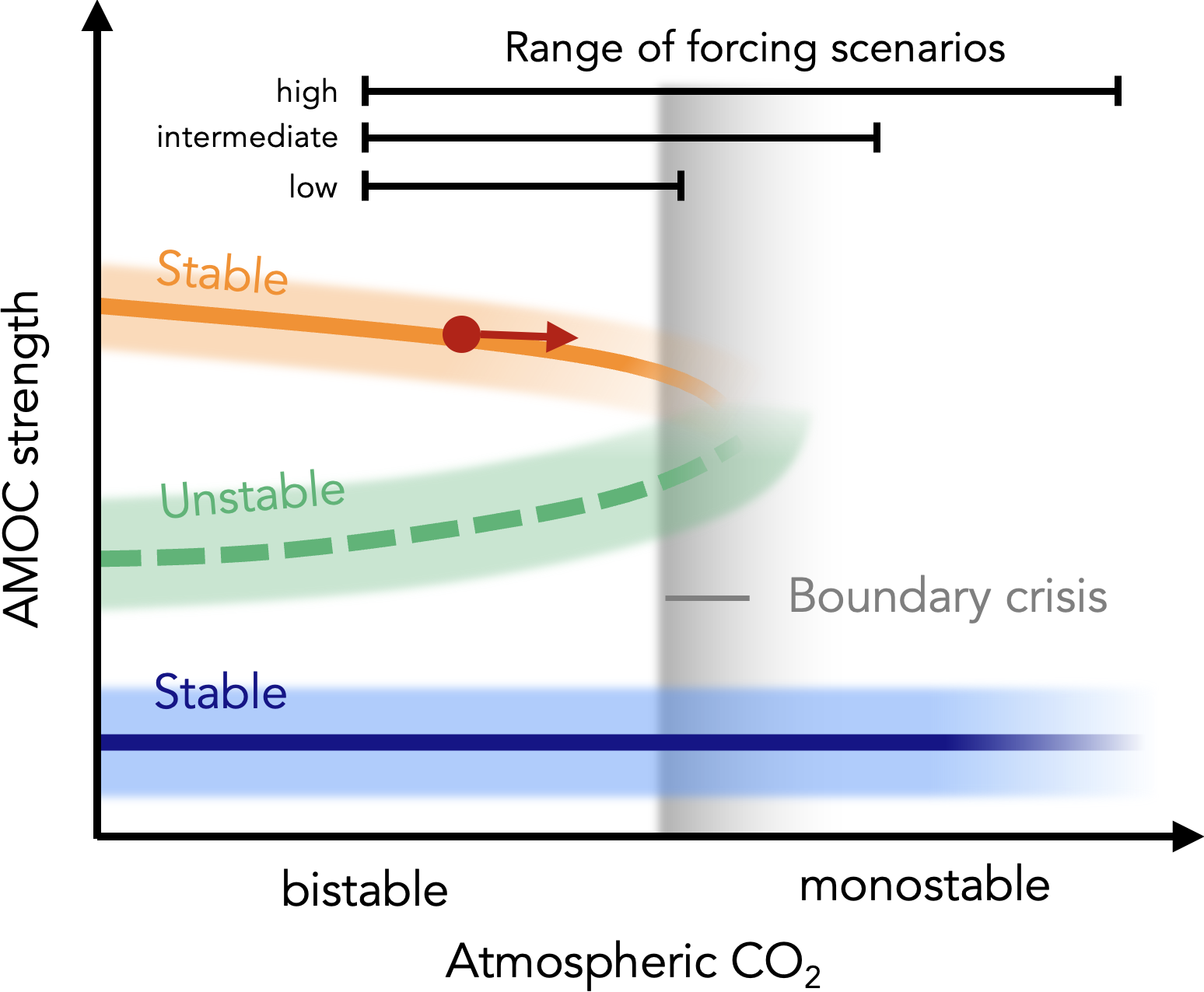}
    \caption{\label{fig:schematic} \textbf{Schematic of the stability setting proposed in this study.} We investigate the global stability of the AMOC at two CO$_2$ levels, one in a bistable regime, where an unstable edge state (green) separates the stable AMOC states, and one in a proposed monostable regime, near a boundary crisis. We then use the results to understand the AMOC behaviour under time-dependent CO$_2$ forcing scenarios (black ranges). The gray region indicates the tipping window in which long transients and ensemble splitting may occur.}
\end{SCfigure}

In this work, we take a global view on the state space of the climate system.
We explore the stability landscape of PlaSim-LSG, an Earth system model of intermediate complexity featuring a bistable AMOC under present-day conditions. 
Instead of focusing on the climates of the stable AMOC states, we investigate the edge state that lies in between and its role for transient dynamics.
Building on recent studies using a conceptual climate model \cite{mehling_limits_2024} and a global ocean circulation model \cite{lohmann_melancholia_2024}, the key novelty here is that we consider a fully coupled modelling setup with a simplified yet Earth-like description of the ocean, atmosphere, cryosphere, hydrosphere and their interactions.

We perform edge tracking at three different CO$_2$ concentrations representing the preindustrial, present-day and possible future climate. In the bistable regime (see Fig. \ref{fig:schematic}), the edge state found in the model exhibits strong AMOC oscillations on centennial timescales, revealing a much richer dynamics than seen in Ref. \cite{lohmann_melancholia_2024}, driven by atmosphere-ice-ocean interactions. Combining simulations under autonomous and nonautonomous forcing, we demonstrate that in our model the AMOC undergoes a boundary crisis at CO$_2$ levels exceeded even under intermediate emission scenarios proposed by the IPCC. At the crisis, the ON state merges with the edge state giving rise to a so-called ghost state, a long-lived yet unstable chaotic set \cite{feudel_multistability_2018,mehling_limits_2024,koch_ghost_2024}. Near but beyond the crisis, we observe centennial to millennial transient behaviour which alternates between modes of variability reminiscent of the ON and edge states, before the circulation ultimately approaches the OFF state.
Our findings help explaining the key aspects of the divergent AMOC behaviour observed in more comprehensive earth system models \cite{romanou_stochastic_2023}.

Our paper is structured as follows. After introducing the model, we describe its AMOC bistability for the present-day climate and assess the AMOC response to transient CO$_2$ forcing until the year 3000 CE (section \ref{sec:model}). In section \ref{sec:climate}, we implement the edge tracking algorithm to construct an edge state of the AMOC (section \ref{sec:global}) and characterise its dynamical and physical properties . In section \ref{sec:crisis}, we explore how the stability landscape changes as a function of CO$_2$ level, revealing a boundary crisis. Relating this to transient simulations in a reduced state space allows to interpret the dynamics observed under time-dependent CO$_2$ forcing (section \ref{sec:ssp-forcing}, where we directly compare with the GISS simulations \cite{romanou_stochastic_2023}). 

\section{AMOC stability in PlaSim-LSG \label{sec:model}}

PlaSim-LSG\footnote{The PlaSim-LSG model is available as open-source code at \url{https://github.com/jhardenberg/PLASIM}.}, the climate model used in this study, is a coupled general circulation model of intermediate complexity, comprising a dynamic ocean, atmosphere, sea ice component and hydrological cycle \cite{fraedrich_planet_2005,maier-reimer_mean_1993,angeloni_evaluation_2020}. Ice sheets and vegetation are prescribed in our setup. 
With around $10^5$ degrees of freedom, the model offers a middle ground between reduced-order models and earth system models \cite{claussen_earth_2002}, producing around 700 model simulation years per day on a single CPU.

Versions of the model have previously been used to study its climate variability \cite{angeloni_climate_2022}, optimal fingerprinting of climate change \cite{lucarini_detecting_2024}, the Snowball Earth transition \cite{kaszas_snowball_2019,margazoglou_dynamical_2021}, and extremes \cite{derrico_present_2022}, particularly using rare event simulations \cite{ragone_computation_2018,wouters_rare_2023,sauer_extremes_2024,cini_simulating_2024}. PlaSim-LSG has also been employed for investigating the multicentennial variability \cite{mehling_high-latitude_2022} and spontaneous tipping \cite{cini_simulating_2024} of the AMOC.

\subsection{Model configuration}

The atmosphere component of the Planet Simulator (PlaSim) \cite{fraedrich_planet_2005} solves the moist primitive equations, describing the conservation of mass and momentum as well as basic thermodynamics, using simplified parameterisations of radiation, convection, precipitation and cloud processes. The prognostic equations are formulated in a spectral represenation truncated at T21 resolution horizontally (roughly giving a $5.6^\circ\times5.6^\circ$ grid) with 10 vertical levels.
The atmosphere is coupled to the Large Scale Geostrophic (LSG) ocean model \cite{maier-reimer_mean_1993}, whereby the 50\,m-thick uppermost vertical layer of LSG is used to compute air-sea fluxes. Assuming that the nonlinear terms of the Navier-Stokes equations can be neglected for large-scale ocean flows \cite{hasselmann_ocean_1982}, the model solves the equations for momentum, temperature and salinity based on hydrostatic balance and the Boussinesq approximation. Convection is not explicitly resolved but accounted for via a convective adjustment scheme. At each time step, the scheme mixes vertically adjacent grid boxes whenever they are unstably stratified, starting from the top and iterating through the water column. Discretised on an E grid \cite{arakawa_computational_1977}, LSG has an effective horizontal resolution of $3.5^\circ \times 3.5^\circ$ and 22 vertical layers on a stretched grid with thicknesses ranging from 50\,m at the surface to 1000\,m in the deep ocean.
The thermodynamic sea ice module is based on a zero-layer model \cite{semtner_model_1976} that computes the ice thickness from the thermodynamic balance at the ice-air and ice-ocean interface, accounting for snowfall. Sea ice transport is neglected.

We configure the model to roughly reflect present-day climatic conditions (with orbital parameters corresponding to around 2000 CE). Its climate sensitivity lies just above $4^\circ$C \cite{angeloni_evaluation_2020}, consistent with the CMIP6 range \cite{zelinka_causes_2020}. 
At the baseline atmospheric CO$_2$ concentration of 360\,ppm (a level recorded in 1995), the default initialization of the model produces an AMOC with a volume transport of around 16\,Sv at 26$^\circ$N ($1\,\text{Sv} = 10^6$\,m$^3$\,s$^{-1}$), close to today's observed value of $16.9\pm 1.2$\,Sv \cite{johns_towards_2023}. Here the AMOC strength is defined as the maximum of the Atlantic meridional streamfunction $\Psi$ at a given latitude $\phi$, taken over the depth $z$ (below sea level), with
\begin{align} \label{eq:streamfunction}
    \Psi(\phi, z,t) = - \int_{z_0}^{z} \int_{\varphi_W}^{\varphi_E} v(\varphi, \phi, z',t) \, r_\circ \cos \phi \text{d}\varphi \text{d} z' \,,
\end{align} 
where $v$ is the meridional velocity field, $\varphi$ the longitude (ranging from the western to the eastern boundary of the Atlantic basin, $\varphi_W$ and $\varphi_E$, respectively); $z_0 \geq z$ is the depth of the sea floor and $r_\circ$ denotes Earth's radius. In this study, unless specified otherwise, we take the streamfunction maximum in the latitude band 46-66$^\circ$N, following Refs. \cite{mehling_high-latitude_2022,cini_simulating_2024}.

\begin{SCfigure}
    \includegraphics[width=0.48\columnwidth]{"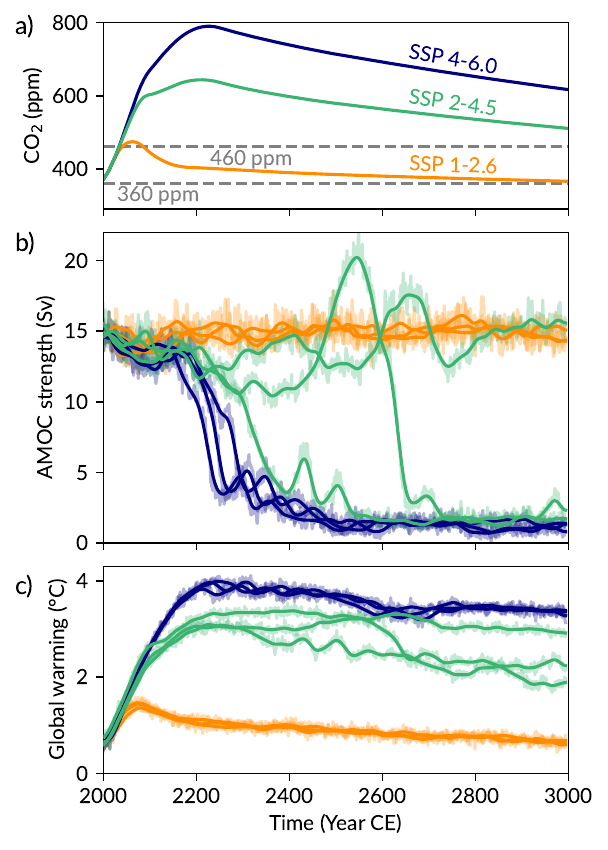"}
    \caption{\label{fig:ssp-triad} Simulated evolution of the AMOC and global warming in PlaSim-LSG under three extended SSP scenarios from 2000 to 3000 CE. a) Atmospheric CO$_2$ concentration for each scenario, indicating 360 and 460\,ppm as dashed lines. b) AMOC strength (10-year smoothed, with annual variability shown as faint lines) for simulations (three ensemble members each) forced by the corresponding SSP scenario as color-coded. c) as b) but showing global mean surface temperature change relative to the 1850-1900 reference.}
\end{SCfigure}
\subsection{Transient CO$_2$ forcing experiments}

As a motivating experiment, we force the model with CO$_2$ projections of Shared Socioeconomic Pathways (SSPs), standardised scenarios for greenhouse gas concentrations until 2500 \cite{meinshausen_shared_2020} (Fig. \ref{fig:ssp-triad}a). For each SSP, we launch an ensemble of simulations (10 members) starting in the forcing year 1995, with initial conditions branched off from a 2000-year control run at 360\,ppm. We compare low emissions (SSP1-2.6), intermediate emissions (SSP2-4.5), and high emissions (SSP4-6.0). Beyond 2500, we assume that the CO$_2$ concentration decays exponentially to 330\,ppm at the rate reached in the decade before 2500.

The AMOC shows qualitatively different behaviour under the three climate change scenarios (Fig \ref{fig:ssp-triad}b). For SSP1-2.6, the vigorous AMOC state is maintained over the 1000-year simulation, as exemplified for three ensemble members. In the SSP4-6.0 scenario, the AMOC collapses in the North Atlantic for all ensemble members. The abrupt decline starts after 2100 and happens within a century. Strikingly, in the intermediate SSP2-4.5 scenario, the ensemble splits, with the AMOC at 46-66$^\circ$N sometimes persisting and sometimes collapsing after strongly varying transients. Even though all ensemble members experience an identical time-dependent forcing, the internal variability leads to a qualitatively differing AMOC response. This difference imprints itself on the global climate, including the global mean surface temperature (Fig. \ref{fig:ssp-triad}c). Global warming under SSP2-4.5 can differ by up to 1$^\circ$C depending on the state of the AMOC. Generally, AMOC weakening reduces the global mean surface temperature, in line with expectations \cite{orbe_atmospheric_2023}.

The results shown in Fig. \ref{fig:ssp-triad} should not be taken as reliable future climate projections, given the reduced complexity of the model, biases \cite{angeloni_evaluation_2020}, and the fact that we neglect other climate-relevant forcings besides CO$_2$, such as methane and aerosol emissions or land-use change. Nonetheless, the AMOC behaviour under SSP2-4.5 is reminiscent of the ensemble splitting found in the more comprehensive GISS model \cite{romanou_stochastic_2023} under the same scenario, as we discuss in section \ref{sec:giss}.

\subsection{Bistability of the AMOC}
\begin{figure*}
    \includegraphics[width=\textwidth]{"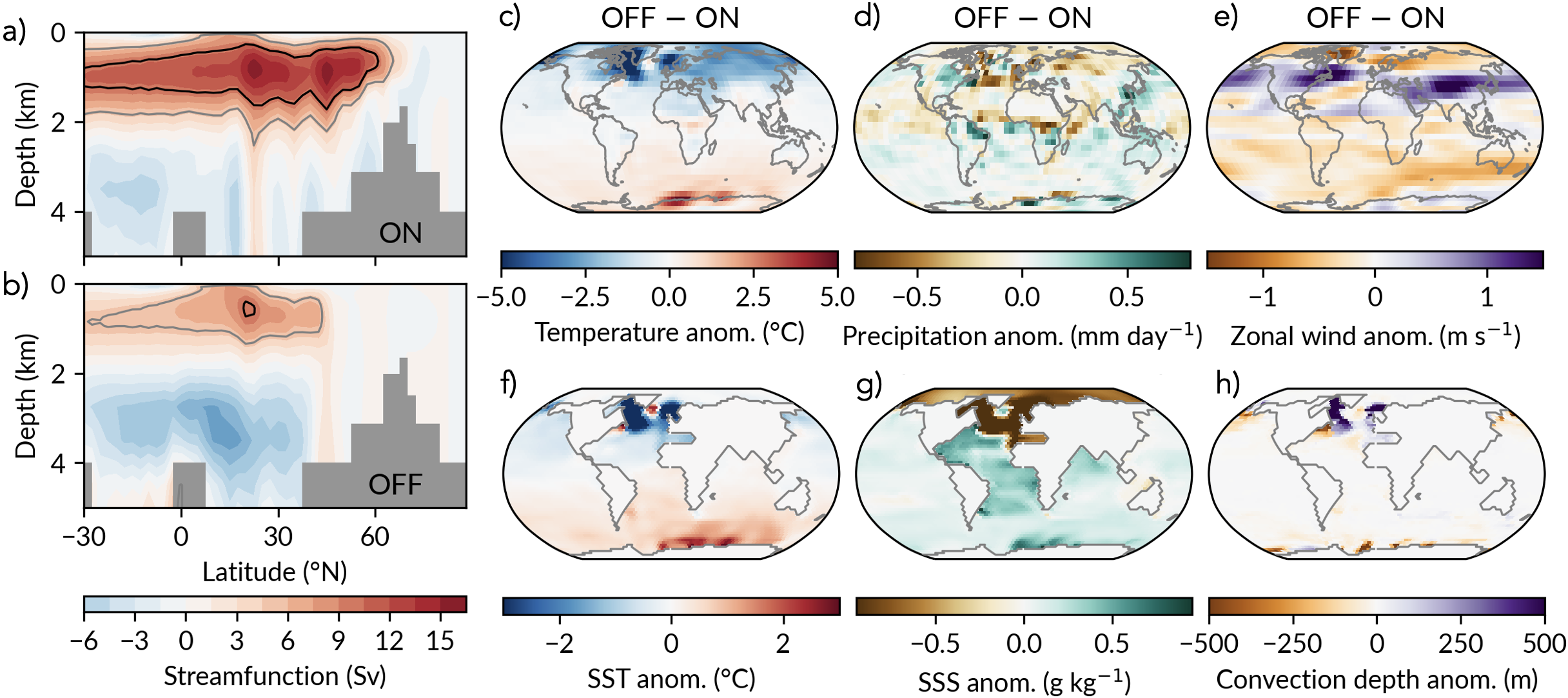"}
    \caption{\label{fig:onof360} \textbf{AMOC bistability in PlaSim-LSG} at  CO$_2$. (a) and (b) show the Atlantic meridional overturning streamfunction for ON and OFF, respectively. (c)-(h) show anomalies of OFF relative to ON for (c) surface air temperature, (d) precipitation, (e) zonal wind speed in the mid-troposphere (around 300-800\,hPa), (f) sea surface temperature, (g) sea surface salinity, and (h) oceanic convection depth. All panels are computed from 1000-year time averages.}
\end{figure*}

At 360\,ppm CO$_2$, the model features (at least) two distinct stable AMOC states: a strong overturning cell with an average strength of 16\,Sv (ON state) and a much weaker and shallower overturning circulation that shuts down to less than about 2\,Sv north of 46$^\circ$N (OFF, Figs. \ref{fig:onof360}a, b). Their stability has been verified via 4000-year long unforced simulations. Determining the precise CO$_2$ range of the bistable regime is challenging due to the occurrence of long transients, as we discuss below.

The ON state resembles the present-day climate and large-scale ocean circulation \cite{buckley_observations_2016}. In the OFF state, the Atlantic meridional streamfunction collapses in the region of the Subpolar Gyre (SPG), while a weakened overturning remains at lower latitudes ($\approx 8\,Sv$ at 26$^\circ$N, see Fig. \ref{fig:onof360}b). Thus, the OFF state in PlaSim-LSG represents a weak, rather than fully collapsed, AMOC. A weak stable AMOC state is found in some models \cite{jackson_understanding_2023, romanou_stochastic_2023}, while in other models the OFF state corresponds to a full collapse of the streamfunction (e.g., \cite{van_westen_asymmetry_2023,lohmann_multistability_2024}).

Still, the OFF state is characterised by the typical climate signal associated with an AMOC collapse, including a reduction of mean surface air temperature in the Northern Hemisphere (locally exceeding 10$^\circ$C), a drying of North Atlantic regions including northern Europe, and a southward shift of the tropical rain belt (Intertropical Convergence Zone) (Fig. \ref{fig:onof360}) \cite{liu_overlooked_2017,bellomo_future_2021}. We also find a strengthened polar jet stream in the northern hemisphere, combined with a large-scale reduction of zonal winds in other regions.

The time-averaged sea surface temperature (SST) is more than 2$^\circ$C (up to 9$^\circ$C) colder compared to the ON state in large parts of the North Atlantic, while the Southern Ocean is up to 3$^\circ$C warmer. The Atlantic subtropical gyre, southern Atlantic, Indian Ocean and Southern Ocean are saltier in the OFF state, whereas the North Atlantic and Arctic Ocean are substantially fresher (except the Irminger Sea). This is a clear signature of the salt-advection feedback and meridional ocean heat transport: the weakened AMOC transports less salinity and heat from the tropics to the north.

The AMOC is closely connected with sites of deep oceanic convection in the North Atlantic, where dense water sinks. In models and observations, major deep convection sites are located in the Labrador Sea (LabS), Irminger Sea (IrmS), and Norwegian Sea (NorS; see Fig. S1 of the Supplemental Information for a map). In PlaSim-LSG, the transition from the ON to the OFF state is characterised by a shutdown of deep convection in the LabS and NorS (Fig. \ref{fig:onof360}f), while the convection depth (as defined in the Supplemental Information) increases in several other locations.

In summary, the AMOC ON and OFF states have a qualitatively different climate on a global scale. To understand the transition behaviour between these states, we now investigate the global stability of the AMOC beyond the steady states.

\section{Beyond stable states: A global view}\label{sec:global}\label{ref:stability}
Generally, a climate model may be viewed as a nonautonomous dynamical system, where the climate state $\bm x(t) \in \mathbb{R}^D$ evolves over time $t \geq 0$ according to
\begin{align} \label{eq:ode}
    \frac{\text{d}\bm x}{\text{d}t} = \bm F\big(\bm x, \, \bm \Lambda(t)\big) \,, \quad \bm x(0) = \bm x_0 \,.
\end{align}
Here $\bm F : \mathbb{R}^D \times \mathbb{R}^K \to \mathbb{R}^D$ may depend explicitly on time via the $K$-dimensional external forcing input $\bm \Lambda : \mathbb{R} \to \mathbb{R}^K$ \cite{wieczorek_rate-induced_2023}. For fixed external forcing $\bm \Lambda(t) = \bm \lambda$, the dynamics is given by the so-called frozen system $\dot{\bm x} = \bm F(\bm x, \, \bm \lambda)$ \cite{wieczorek_rate-induced_2023}. Multistability is characterised by the coexistence of multiple attractors for given $\bm \lambda$. In our case, with $\bm \lambda = \lambda_{\text{CO}_2}=360\,$ppm, there are two chaotic attractors corresponding to the ON and OFF states. Each attractor possesses a basin of attraction, i.e., a set of initial conditions $\{ \bm x_0 \}$ that evolve towards it as $t\to\infty$. Since PlaSim-LSG is fully deterministic, the asymptotic state is thus uniquely determined by the initial condition in the absence of perturbations\footnote{More generally, stochastic parameterisations of unresolved processes change this picture and require a stochastic description}.

The two basins of attraction must be separated by a basin boundary of dimension $D - 1 \leq D_b < D$ with respect to the dimension $D$ (number of degrees of freedom) of the system. The basin boundary between the two AMOC states in PlaSim-LSG is thus a high-dimensional set in the model state space. As was shown for conceptual \cite{mehling_limits_2024} and intermediate-complexity \cite{lucarini_edge_2017} climate models, the basin boundary can be fractal with almost full state space dimension ($D_b\ll1$).
The basin boundary is crucial in the context of critical transitions because it marks the threshold in state space where the dominant feedback changes from a stabilizing (negative) feedback that exerts a restoring force towards the original attractor to a destabilizing (positive) feedback that drives a self-perpetuating transition to a competing attractor.

Even if the \enquote{curse of dimensionality} \cite{altman_curses_2018} currently prevents computing the quasipotential of high-dimensional systems, we can learn much about its structure by analysing edge states as important landmarks therein. Since edge states are unstable, they generally cannot be found by direct simulation or basic continuation. However, since they are unstable in only one direction (transversal to the basin boundary), edge states can nonetheless be computed purely based on forward integration of the model, as described in the following. 

\subsection{Finding the basin boundary}

How can we locate the basin boundary between two competing AMOC states in a high-dimensional climate model? We need a pair of initial conditions $\bm x_a$ and $\bm x_b$ that are attracted by the ON and OFF state, respectively. Interpolating along a straight line in state space between $\bm x_a$ and $\bm x_b$ necessarily leads to crossing a segment of the basin boundary lying somewhere in between -- provided that the interpolation is performed in the full state space including all degrees of freedom (prognostic variables).

Here, we take two model restart files from previous PlaSim-LSG simulations \cite{mehling_high-latitude_2022} as initial conditions $\bm x_a$ and $\bm x_b$. The simulations were performed at 285\,ppm CO$_2$ with differing vertical diffusivity profiles, which strongly affected the AMOC strength \cite{mehling_high-latitude_2022}. In our model configuration (at 360\,ppm, see Supplemental Information for details), these two initial conditions are located near the ON and OFF state, respectively, and evolve to these.

Now, interpolating along all variables $x_i$ with $i = 1, \dots, D$ in the model restart files ($D \approx 10^5$), we compute new initial conditions $x_{j, i} = x_{a,i} + 0.1j( x_{b,i} - x_{a,i})$ for $j=1,2,\dots,9$.
Computing the meridional streamfunctions (Eq. \eqref{eq:streamfunction}) for these initial conditions shows that the states $\bm x_j$ are monotonically decreasing in AMOC strength with increasing $j$ (Fig. \ref{fig:edgetrA2_result}a). The differences in AMOC strength between adjacent states are not equidistant, reflecting that the AMOC strength is a nonlinear mapping of the full state space.

From the initial conditions $\bm x_j$, we run parallel simulations for 200 years each. While the trajectory initialised at $\bm x_1$ remains close to the ON state in AMOC strength, all other trajectories lead to the OFF state. This implies that a part of the basin boundary is located between $\bm x_1$ and $\bm x_2$ in state space. This pair of initial conditions constitutes the starting points of our edge tracking procedure.

\subsection{Edge tracking algorithm \label{sec:edgetracking}}
\begin{figure}
    \centering
    \includegraphics[width=0.75\columnwidth]{"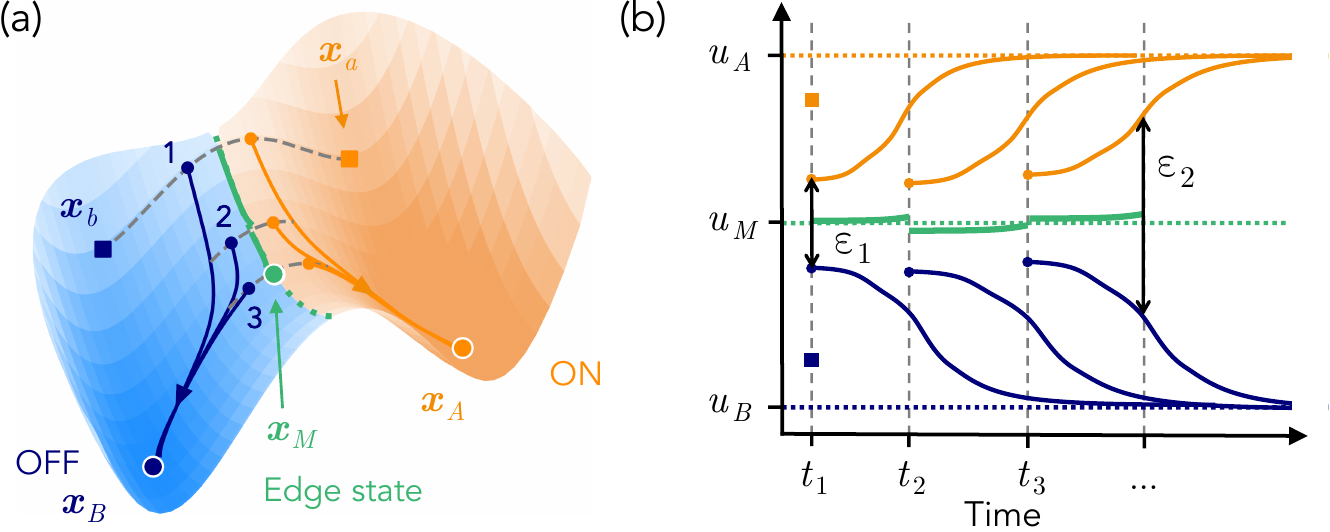"}
    \caption{\label{fig:edgetracking_scheme} \textbf{Edge tracking method}, illustrated (a) in the schematic quasipotential landscape of a bistable AMOC and (b) as idealised timeseries projected onto the coordinate $u$ (in our context, the AMOC strength). The landscape shows the basins of attraction of the attractors $\bm x_A$ (orange shading) and $\bm x_B$ (blue shading), separated by the basin boundary (green dashed). Starting from $\bm x_a$ and $\bm x_b$, three exemplary iterations (as numbered) yield a pseudotrajectory (green solid line) that leads close to the edge state $\bm x_M$. Gray dashed lines indicate the bisections. 
    }
\end{figure}

The edge tracking algorithm, as originally proposed in Refs. \cite{battelino_multiple_1988,skufca_edge_2006} and adopted in Refs. \cite{lucarini_edge_2017,mehling_limits_2024,lohmann_melancholia_2024,borner_climate_2025},
consists of an iterative loop with two steps:
\begin{enumerate}
    \item \textit{Bisection.} Between two initial conditions converging 
    to attractors $A$ and $B$, respectively, bisect repeatedly along a straight line in state space to obtain two new initial conditions that are less than a distance $\varepsilon_1$ apart while still converging to different attractors (one to $A$ and the other to $B$).
    \item \textit{Tracking.} From each of the two new initial conditions, run a simulation in parallel. Stop the simulations when the two trajectories diverge by more than a distance $\varepsilon_2$, and use the end points of these simulations as initial conditions for the next iteration. Repeat 1.
\end{enumerate}

Here the distance measures $\varepsilon_{1,2}$ could be the Euclidean distance in a normalised state space or any other appropriate measure of separation between the two states. We simply measure the difference in 10-year smoothed AMOC strength.

Running the algorithm yields two series of trajectory segments that shadow the basin boundary on either side of it. By concatenating the segments and averaging over both series at each time point, we obtain a pseudotrajectory that approximates a trajectory on the basin boundary. The repeated rebisection of initial conditions thereby counteracts the instability that causes any trajectory initialised near the boundary to eventually diverge from it. Based on the property of edge states as attracting sets when restricted to the boundary, the pseudotrajectory is expected to converge to an edge state. If this state supports chaos, the algorithm converges to this chaotic set and successively populates its invariant measure.

The algorithm is computationally expensive in complex models, especially when trajectories converge slowly to the attractors. This is because the asymptotic state of each new initial condition must be determined by simulation, which can take hundreds of model years for the AMOC. With PlaSim-LSG, however, we can exploit the fact that multiple simulations can be run in parallel. Thus, instead of successive bisections as described in step 1 above (and implemented by, e.g., Refs. \cite{mehling_limits_2024,lohmann_melancholia_2024}), we compute nine equidistant initial conditions $\bm x_j^{(k)}$ at once by linear interpolation and run parallel simulations from them. That way, we can reduce the distance $\varepsilon$ between initial conditions by a factor ten in one interpolation step $k$. A pseudocode detailing our implementation is provided in the Supplemental Information.

\subsection{Converging to an edge state}

\begin{figure*}
    \centering
    \includegraphics[width=0.95\textwidth]{"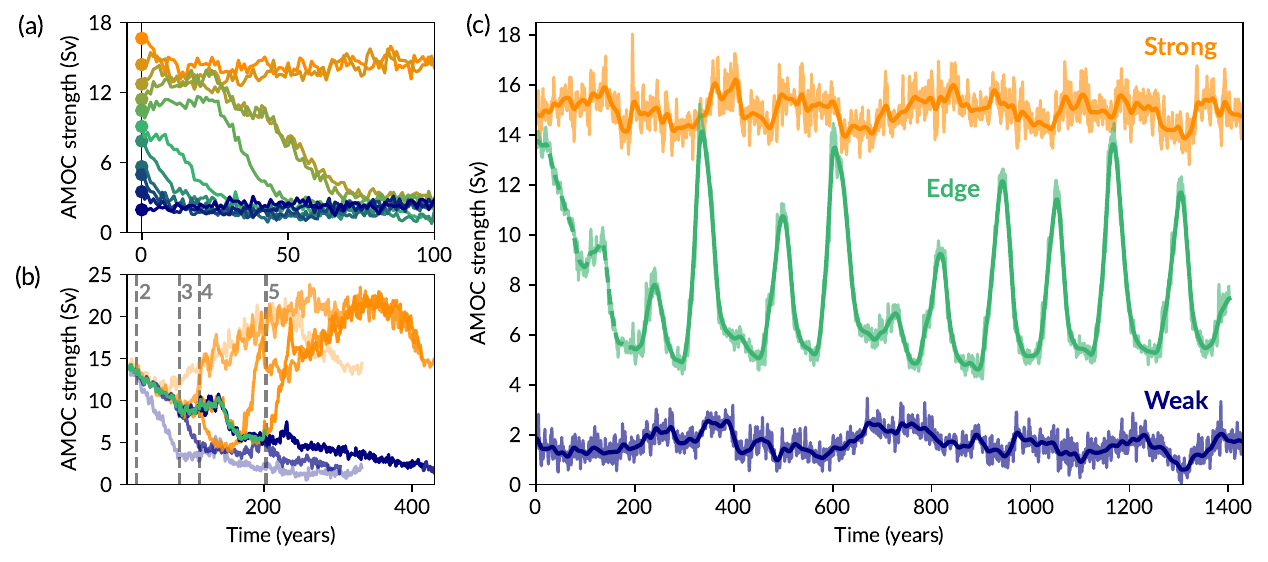"}
    \caption{\label{fig:edgetrA2_result} \textbf{Edge tracking and AMOC states} at 360ppm CO$_2$. (a) Interpolating initial conditions between the ON (orange) and OFF (blue) AMOC state allows locating the basin boundary. (b) Iterations 2-5 of the edge tracking algorithm, showing the trajectories that converge to ON (orange) and OFF (blue), respectively. The edge pseudotrajectory (green) is constructed from segments of these trajectories. (c) Edge trajectory (green) and trajectories on the ON (orange) and OFF (blue) attractors. The AMOC strength is measured between 46-66$^\circ$N.}
\end{figure*}

For the first few iterations of edge tracking, spanning about 200 years, the resulting pseudotrajectory (hereafter called \textit{edge trajectory}) decreases in AMOC strength from 14\,Sv to about 5\,Sv. Subsequently, the edge trajectory begins a series of large AMOC oscillations (Fig. \ref{fig:edgetrA2_result}c). The quasiperiodic oscillations vary in amplitude from 3 to 10\,Sv, with a mean period of $118 \pm 7$ years (estimated from 10 peaks). This behaviour persists until the edge tracking was stopped after around 1400 years (39 iterations).

The recurrent pattern of centennial AMOC cycles suggests that the edge trajectory has converged to an edge state and thereafter evolves on this unstable set. This claim is supported by the fact that the specific potential energy of the global ocean is relatively constant after convergence (Fig. \ref{fig:energetics}b) and that the salinity in the deep Pacific, Indian, and Southern Oceans has equilibrated (not shown). Since the oscillations are neither perfectly periodic nor constant in amplitude, the edge state appears to be a chaotic saddle with a more complex geometry compared to an unstable limit cycle. This nonattracting invariant set is approximated by the edge trajectory after removing the initial transient of 200 years.

We emphasise that the edge trajectory varying in time does not mean the edge state itself is time-dependent: since we fix the external forcing, the edge state is invariant in time, and the the edge trajectory reflects the dynamics \textit{on} the edge state.

\subsection{Reduced state space \label{sec:reduced-phase-space}}

\begin{SCfigure}
    \centering
    \includegraphics[width=0.5\columnwidth]{"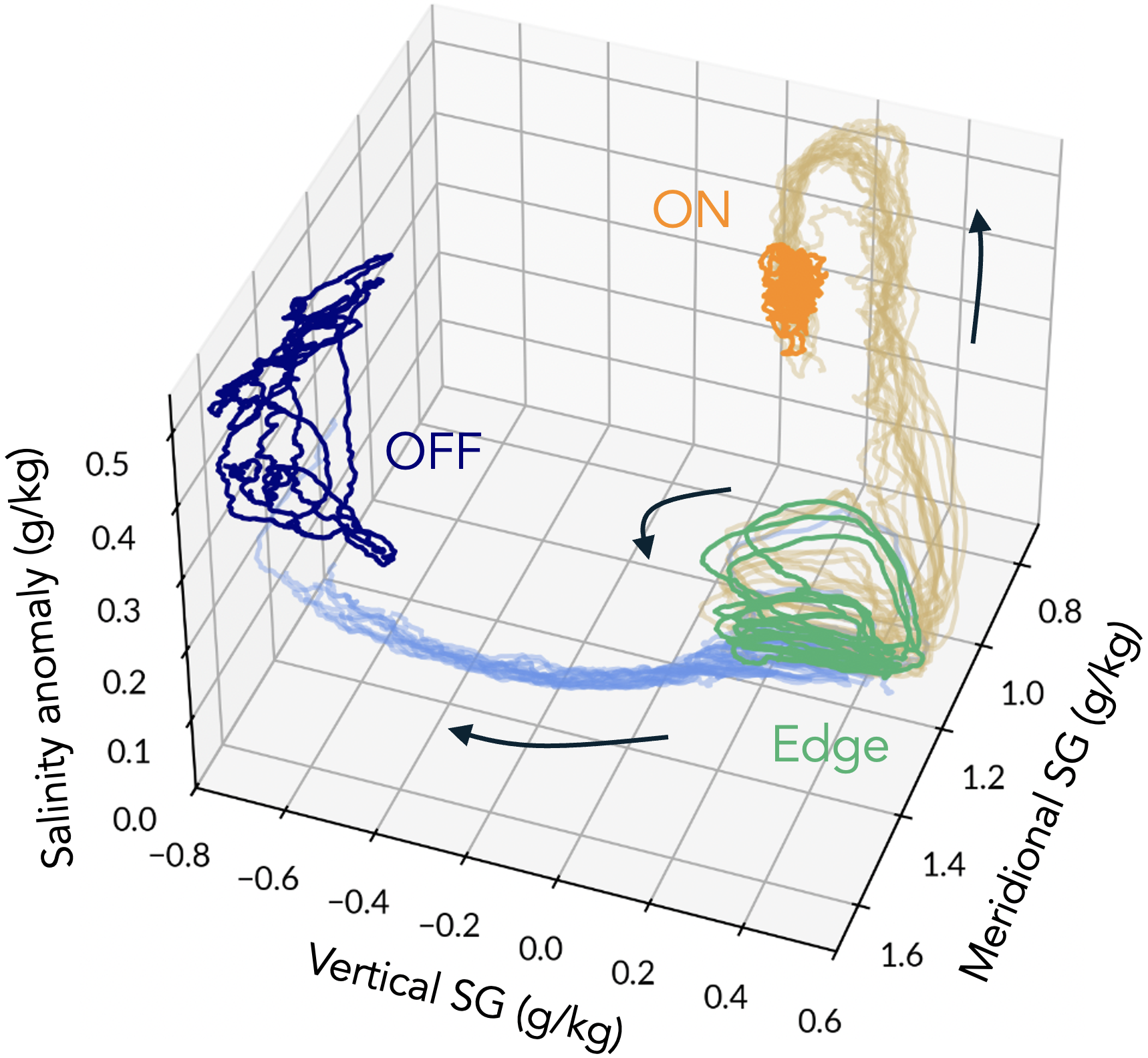"}
    \caption{\label{fig:states3d} \textbf{State space projection} onto the meridional SG, vertical SG, and salinity anomaly in the deep North Atlantic (below 1000\,m, north of $50^\circ$S). Faint orange (blue) lines show trajectories relaxing from near the edge state to the ON (OFF) state. Arrows indicate the time direction.}
\end{SCfigure}

Looking at the one-dimensional AMOC timeseries (Fig. \ref{fig:edgetrA2_result}c) gives the impression that the edge trajectory oscillates back and forth between the ON and OFF states. However, visualizing the dynamics in a reduced state space clarifies that the edge state is separated from the attractors (Fig. \ref{fig:states3d}).

Determining a suitable low-dimensional projection of the $10^5$-dimensional dynamics is challenging due to the countless possible combinations of variables. Based on our physical understanding of the AMOC, we consider the zonally averaged salinity field in the Atlantic. An empirical orthogonal function (EOF) analysis \cite{navarra_guide_2010} combining 20\,000 years of edge tracking simulations shows that this field contains sufficient information to disentangle the dynamics (Figs. S4 and S5 of the Supplemental Information). Specifically, the two leading EOFs reveal that most of the variance is explained by a meridional salinity dipole in the upper 1000\,m and a vertical dipole in the North Atlantic. From this we derive a reduced state space spanned by the following three variables:
\begin{itemize}
    \item The \textit{meridional} salinity gradient (SG) in the Atlantic, measured as the mean salinity difference between 0-20 N and 40-80 N in the top 1000\,m (omitting the top 100\,m),
    \item The \textit{vertical} SG in the North Atlantic, defined as the mean salinity difference between the depths 100-1000\,m and 1000-3000\,m at 46-66$^\circ$N,
    \item The \textit{deep North Atlantic salinity anomaly}, defined as the mean salinity anomaly relative to 35\,g\,kg$^{-1}$ in the Atlantic basin north of 50$^\circ$N and below 1000\,m depth.
\end{itemize}
The benefit of using these variables, instead of directly using the principal components of the EOFs, is that they can easily be computed for any spatially resolved ocean model, permitting inter-model comparisons. The meridional SG is negatively correlated with the AMOC strength, since a stronger AMOC transports more salt to the North Atlantic, reducing the salinity difference between low and high latitudes. The vertical SG and deep salinity anomaly are related to deep convection and the stability of the water column in the North Atlantic.

Viewing the trajectories of Fig. \ref{fig:edgetrA2_result} in the reduced state space, we see that each of the ON, OFF, and edge states occupies a distinct region (Fig. \ref{fig:states3d}). The edge state has a higher vertical SG and fresher deep North Atlantic than both the ON and OFF states. The OFF state has the saltiest deep North Atlantic and largest meridional SG. While the ON state covers a relatively small volume of the reduced state space, the AMOC oscillations of the edge state are clearly seen as loops in the meridional-vertical SG plane. Also the OFF state exhibits relatively large internal variability that is captured in this projection but not in the AMOC strength. This low-frequency variability on multi-centennial timescales is caused by global inter-basin salt exchanges (not shown).

The simulations used to perform edge tracking also reveal the transition pathways from the edge state to each of the the ON and OFF states. The trajectories of the final 20 edge tracking iterations reveal clear characteristic pathways to either attractor (Fig. \ref{fig:states3d}), which trace the unstable manifold of the chaotic edge state.

\section{Climate of the edge state \label{sec:climate}}
The pseudotrajectory on the edge state is constructed from segments of actual model trajectories, meaning that we can explore its weather and climate as with any other model simulation. This provides insight into what the world looks like near the edge state and into the processes involved in the instability of the AMOC.

\subsection{Energetics}

\begin{figure*}
    \centering
    \includegraphics[width=\textwidth]{"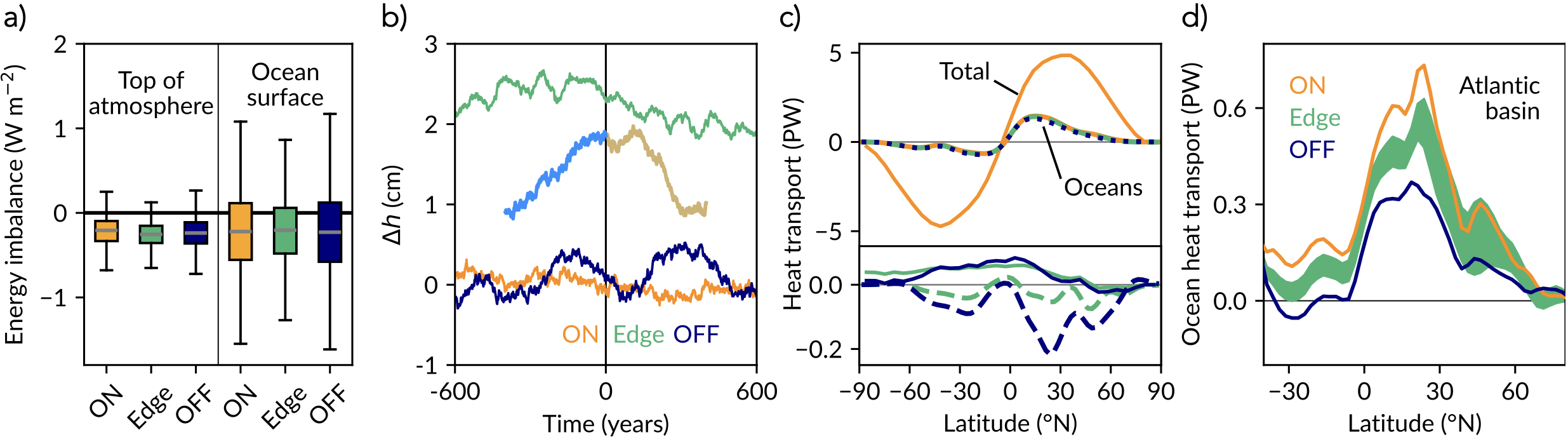"}
    \caption{\label{fig:energetics} \textbf{Energetics of the climate states.} (a) Imbalance of top of the atmosphere radiation (left) and heat flux at the sea surface (right), integrated over the globe for ON, OFF and Edge (negative imbalance means the Earth/ocean is losing energy). (b) Oceanic enter of mass anomaly $\Delta h$ (relative to 1970.3126\,m below sea level) for years 200-1400 of the edge trajectory and corresponding time intervals for ON and OFF. The 400-year long relaxation paths from Edge$\to$ON (beige) and Edge$\to$OFF (light blue, plotted in reverse time) are shown for one of the edge tracking iterations. (c) Northward meridional heat transport, showing the total from atmosphere and oceans (for ON, orange) and the oceanic contribution (all states, dotted). Bottom inset: Difference in total (solid) and oceanic (dashed) heat transport for OFF (blue) and Edge (green) relative to ON. (d) Oceanic heat transport in the Atlantic basin only, showing the variability of the AMOC oscillation on the Edge state (green band).}
\end{figure*}

The first question is whether the edge state energetically fulfills steady state conditions, requiring an approximately vanishing global energy budget for the coupled climate system and its subcomponents \cite{peixoto_physics_1992,lucarini_energetics_2011}.  Indeed, both the radiative balance at the top of the atmosphere as well as the globally integrated net surface heat flux between the ocean and atmosphere are close to zero (comparable to the ON and OFF states; Fig. \ref{fig:energetics}a).

The meridional heat transport of the ocean and atmosphere combined is nearly identical for the ON, OFF and edge states, despite differences in the ocean circulation (\ref{fig:energetics}b). This means that the atmosphere largely compensates for changes in oceanic heat transport \cite{knietzsch_impact_2015}, manifesting the Bjerknes compensation \cite{bjerknes_atlantic_1964,stone_constraints_1978} also reported in previous studies on the AMOC variability and collapse \cite{povea-perez_central_2024, orbe_atmospheric_2023}. 
Because of the AMOC, the Atlantic Ocean is the only ocean basin with a northward oceanic heat transport on both hemispheres, causing an asymmetry of the oceanic meridional heat transport. A reduced AMOC thus decreases this asymmetry, as we observe for the OFF state (Fig. \ref{fig:energetics}c, lower panel). Interestingly, the change in the atmospheric transport slightly overcompensates the reduction in the oceanic transport (Fig. \ref{fig:energetics}c, upper panel). The time-averaged Atlantic meridional heat transport of the edge state lies in between that of the ON and OFF states, though the AMOC oscillations cause temporal variations of more than 0.1\,PW especially in the northern mid-latitudes (Fig. \ref{fig:energetics}d). 

Based on the picture of a double-well stability landscape of the bistable AMOC (see Fig. \ref{fig:edgetracking_scheme}), we expect that the edge state has a higher potential energy than the ON and OFF states. While a full account of potential energy in the coupled climate system requires considering energy exchanges with all subcomponents, we here propose the oceanic centre of mass,
\begin{align}
     h = H - \frac{\int_0^H z \bar\rho(z) \text{d} z}{\int_0^H \bar\rho(z) \text{d} z} \,,
\end{align}
as an approximate energy measure to compare the oceanic specific potential energy among the different AMOC states \cite{rosenthal_center_2025}. Here $H=6000$\,m is the maximum depth of the sea floor, $z$ is the depth coordinate (positive downwards), and $\bar \rho$ is the horizontally integrated density across the ocean ($\bar \rho = 0$ below the sea floor).

The edge state has a significantly higher centre of mass -- and thus specific potential energy -- compared to the two attractors. This aligns with the situation in a global ocean model, where the dynamic enthalpy of the edge state was shown to be elevated \cite{lohmann_melancholia_2024}. In our case, the ON and OFF states have a comparable centre of mass, with the OFF state exhibiting multi-centennial variability in $h$ due to global salt exchanges, as also observed in Fig. \ref{fig:states3d}.

\begin{figure*}
    \centering
    \includegraphics[width=0.9\textwidth]{"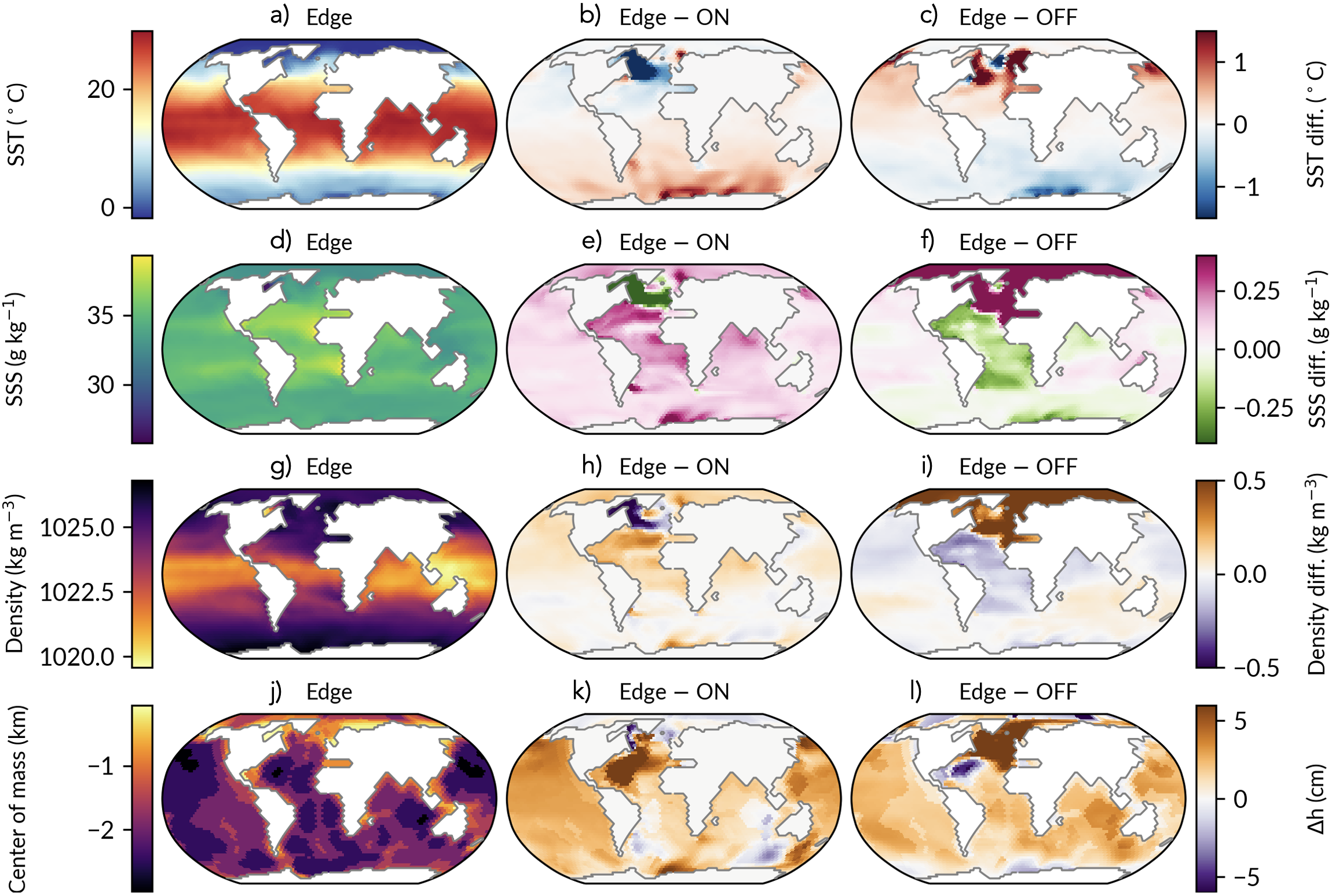"}
    \caption{\label{fig:state-fields} \textbf{Ocean properties of the edge state} (at 360\, ppm CO$_2$), displayed as time averages over the final 640 years of the edge trajectory in absolute values (first row) and as differences relative to the ON (second) and OFF (third row) states: (a)-(c) sea surface temperature (upper 100\,m), (d)-(f) sea surface salinity (upper 100\,m), (g)-(i) surface density (upper 100\,m), and (j)-(l) water column centre of mass. See Fig. S3 (Supplemental Information) for deep sea properties.}
\end{figure*}

To understand which geographical regions contribute most to the higher centre of mass, we calculate the time average of $h$ for the water column at each horizontal grid point. Mapping out the difference $\Delta h$ between the edge state and each of the attractors shows that the edge state has a higher specific potential energy in most of the global ocean, particularly in regions of the North Atlantic (Figs. \ref{fig:state-fields}h-i). Yet, some regions also have a negative $\Delta h$, e.g. in parts of the LabS relative to the ON state and the Atlantic subtropical gyre when compared to the OFF state.

\subsection{Excursive observables \label{sec:excursional}}

Since the edge state lies on the basin boundary between the ON and OFF states, one might expect that its climate lies somewhere in between that of the ON and OFF states, too. For example, in terms of AMOC strength and Atlantic meridional heat transport the edge state oscillates between the ON and OFF states. At the same time, the edge state features a fresher deep North Atlantic than both attractors, and a higher centre of mass. In a high-dimensional system like our climate model, there may be many directions -- which we term excursive observables -- in which the edge state lies outside of the interval bounded by the two attractors. These directions in state space could be particularly relevant for detecting EWS \cite{lohmann_role_2024} and for evaluating transition probabilities via rare event techniques relying on a score function \cite{jacques-dumas_estimation_2024}. Along transition paths (provided that they pass via the vicinity of the edge state), we expect excursive observables to undergo non-monotonic excursions. Thus, the signal of a transition could initially have the opposite sign of the anticipated change.

In most ocean regions, the time-averaged sea surface salinity (SSS) and sea surface temperature (SST) of the edge state lie in between that of ON and OFF states (Fig. \ref{fig:state-fields}). However, almost the entire Arctic Ocean is saltier and denser in the upper ocean relative to the attractors. Parts of the NorS are warmer than both attractors, and the northwestern Pacific Ocean is warmer and saltier on the edge state. Other excursive observables include the sea ice cover in the IrmS and the surface freshwater flux in the NorS.

\subsection{Drivers of the unstable oscillations}
\begin{SCfigure}
    \centering
    \includegraphics[width=0.52\columnwidth]{"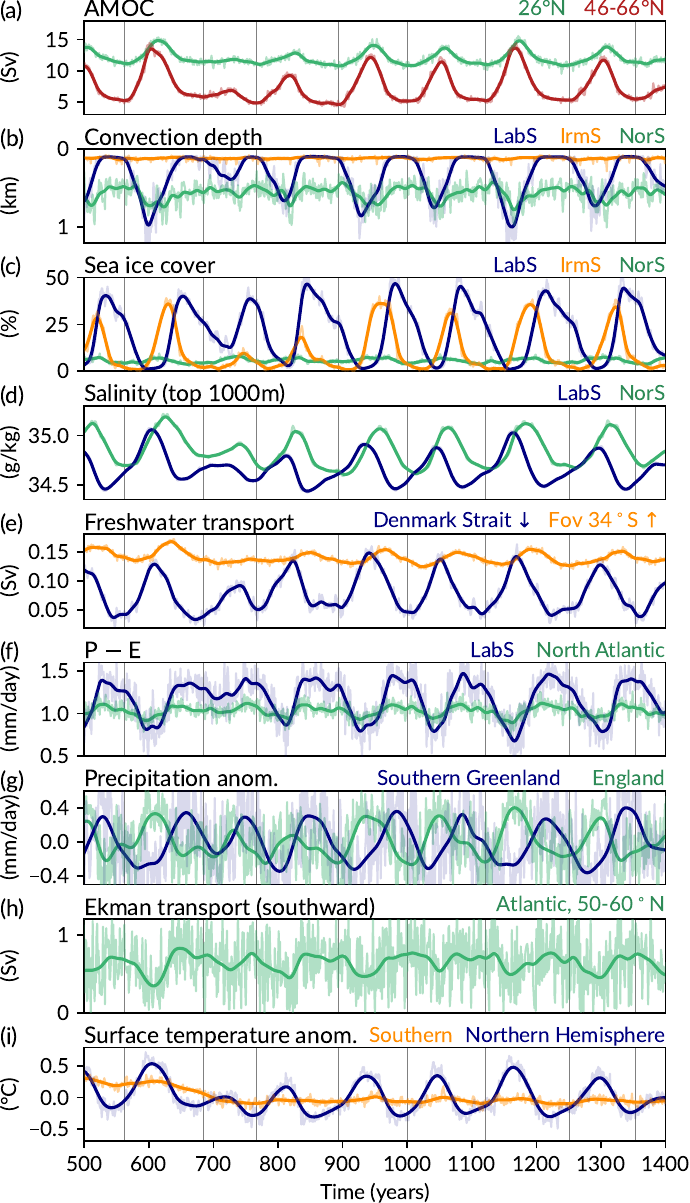"}
    \caption{\label{fig:oscil} \textbf{Oscillations on the edge state} captured in (a) the AMOC strength at 46-66$^\circ$N (red) and 26$^\circ$N (green), (b) the annual-mean maximum convection depth in regions as labeled, (c) sea ice cover in regions as labeled, (d) mean salinity (averaged over the top 1000\,m) in the LabS (blue) and across the Atlantic north of 50$^\circ$N (green), (e) Freshwater transport through the Denmark Strait (southward, blue) and for the overturning component $F_\text{ovS}$ at 34$^\circ$S (orange), (f) LabS (blue) and North Atlantic (green) precipitation minus evaporation, (g) precipitation anomaly over the LabS (blue) and Northern Europe (green), (h) southward Ekman transport in the Atlantic, averaged zonally and over 50-60$^\circ$N, and (i) mean surface temperature anomalies in the Northern (blue) and Southern (orange) Hemispheres. Thick lines are smoothed with a 5-year Gaussian filter (10 years for precipitation and Ekman transport). See Fig. S2 (Supplemental Information) for a lead-lag analysis of these signals.}
\end{SCfigure}

The most prominent dynamical feature of the edge state are the large AMOC oscillations with a period of around 120 years and an amplitude of up to 10\,Sv between 46-66$^\circ$N (Fig. \ref{fig:edgetrA2_result}c). At 26$^\circ$N, the AMOC oscillations are qualitatively similar but have a smaller amplitude (Fig. \ref{fig:oscil}a). Together with the overturning strength, many other climate observables oscillate at this frequency (Fig. \ref{fig:oscil}). What drives the unstable oscillations?

In PlaSim-LSG, the transition from ON to OFF is characterised by a shutdown of all deep convection sites in the LabS, IrmS and NorS (Fig. \ref{fig:onof360}). On the edge state, deep convection persists in the NorS, with some variation linked to the AMOC (Fig. \ref{fig:oscil}b). In the LabS, deep convection undergoes large oscillations, switching on and off in close correspondence with the AMOC strength. Convection is inactive in the IrmS. 

To relate different oscillating variables in time, we compute lag correlations between the AMOC strength at 46-66$^\circ$N and other variables, considering time lags between $-120$ and 120 years. We select variables whose 3-year smoothed timeseries has a maximum lag correlation above 0.8 in absolute value. For these variables, we compute correlation values also for the unfiltered timeseries (annual resolution), giving the values reported in the following and in Fig. S2 (Supplemental Information).

The sea ice cover fraction oscillates strongly in the LabS and, with a phase shift of around $\pi/2$, in the IrmS (Fig. \ref{fig:oscil}c). In both regions, the sea ice retreats almost entirely during the respective minimum.
In the NorS, there is little sea ice on the edge state at all times. A clear phase shift is also seen between the mean upper ocean salinity (top 1000\,m) of the LabS and NorS (Fig. \ref{fig:oscil}d). Generally, the salinity changes could be caused by horizontal advection, convection, or surface freshwater fluxes. We find a strong southward freshwater transport through the Denmark Strait between Greenland and Iceland, oscillating in anti-phase with the AMOC (Fig. \ref{fig:oscil}e), as well as a large amplitude in precipitation minus evaporation (P$-$E) over the LabS (Fig. \ref{fig:oscil}f). Furthermore, the magnitude of the wind-driven southward Ekman transport in the North Atlantic (50-60$^\circ$N) is negatively correlated with the AMOC strength at a lead time of four years (Figs. \ref{fig:oscil}h and S2, Supplemental Information). Around the AMOC minimum on the edge state, the Ekman transport is stronger compared to both the ON and OFF states, indicating a potential role of the wind stress in triggering an overturning decline \cite{cini_simulating_2024} (see Fig. S6 of the Supplemental Information). In the atmosphere, temperature, precipitation and winds likewise display variability on the 120-year timescale (Fig. \ref{fig:oscil}g-i), though correlations with the AMOC strength are lower due to the much higher interannual variability in the atmosphere compared to the ocean.

A key observation is that the upper ocean salinity, deep convection, sea ice, and P$-$E in the LabS all lead the AMOC by 6-8 years, with lag correlations ranging between 0.78 and 0.92 (in absolute value, see Fig. S2, Supplemental Information). P$-$E averaged across the entire North Atlantic (between 50-80$^\circ$N) has an even larger lead time of 11 years, though the correlation is less strong. The AMOC strength measured at 26$^\circ$N follows the AMOC at 46-66$^\circ$N by 6 years, and the overturning component $F_\text{ovS}$ of the freshwater export at the Atlantic southern border (34$^\circ$S) \cite{van_westen_physics-based_2024} lags behind by 26 years. Our analysis thus shows that the LabS is a key region in driving the AMOC oscillations on the edge state. The fact that the salt and volume transport in the tropical Atlantic lags the AMOC strength further north suggests that the salt-advection feedback does not initiate the oscillations, though it likely plays an important role in amplifying them. Rather, ocean-ice-atmosphere interactions in the North Atlantic and Arctic appear crucial for triggering the AMOC cycles.

\begin{figure*}
    \centering
    \includegraphics[width=\textwidth]{"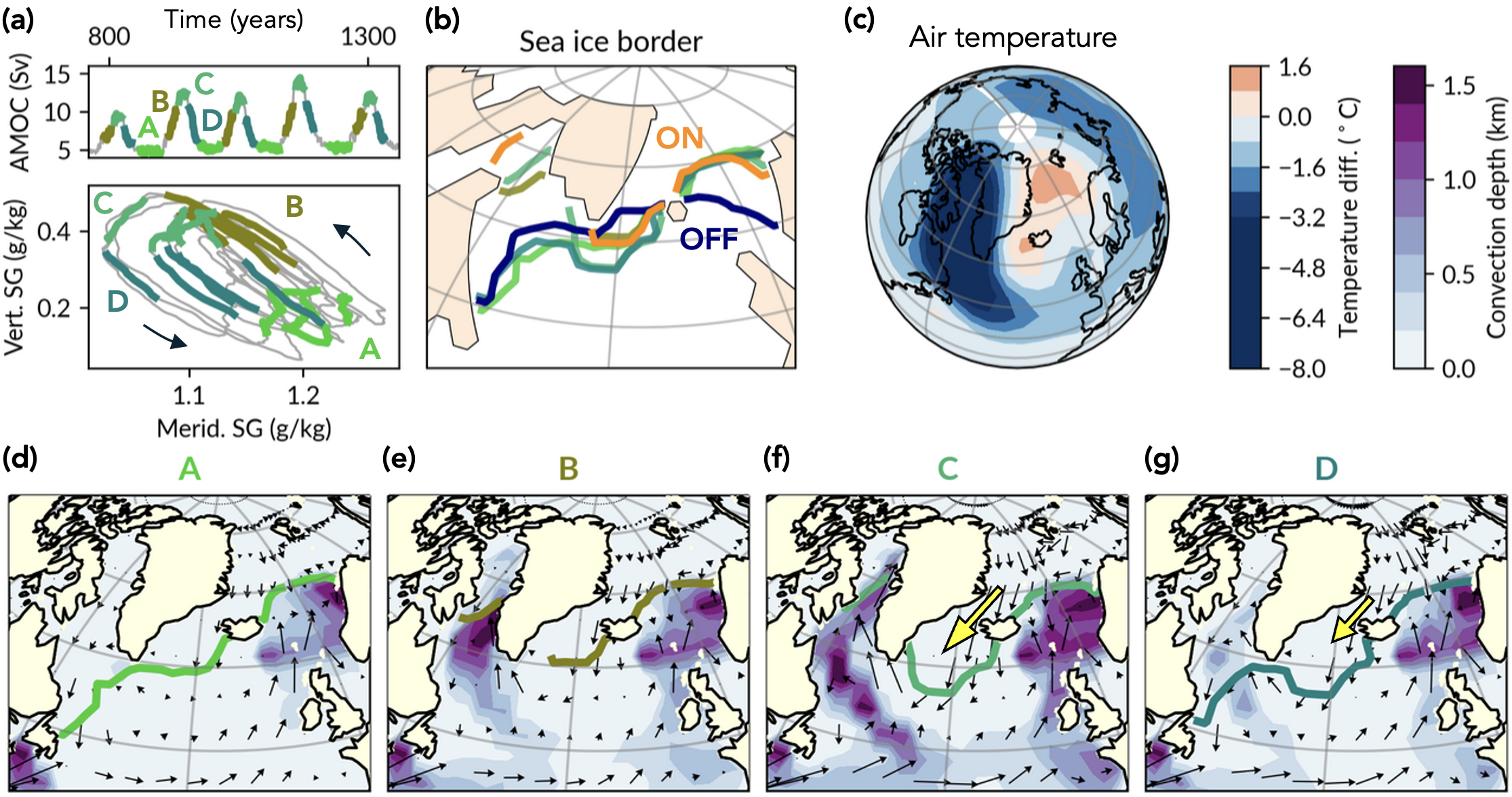"}
    \caption{\label{fig:phases} \textbf{Phases of the AMOC oscillations.} (a) Segmentation of the final 5 oscillations into phases A-D as labeled, shown as a timeseries (top panel) and projected onto the reduced state space of meridional and vertical SG (bottom panel). (b) Annual mean sea ice border for all phases compared with ON and OFF. (c) Surface air temperature difference for A minus C. (d)-(g) Maps of convection depth (shading), surface currents (black arrows) and the sea ice border (thick line) in the North Atlantic for phases A-D, respectively. The yellow arrow represents the strength of the freshwater flux through the Denmark Strait.}
\end{figure*}

To gain further process understanding, we now divide each AMOC cycle into four phases (A--minimum, B--rise, C--maximum, D--decline; see Fig. \ref{fig:phases}) and consider time averages for each phase over the final five oscillations of the edge trajectory. In phase A, the LabS is ice-covered, preventing deep convection and thus maintaining a weak AMOC. Deep water formation in the NorS ensures that the AMOC is not as weak as in the OFF state. In phase B, sea ice retreats in the LabS, allowing the ocean to release heat to the atmosphere and consequently deep convection to be activated. The salt-advection feedback kicks in, supplying warm and salty water to the LabS, enhancing sea ice retreat and convection up to the AMOC maximum in phase C. Then, however, the salinity and convection in the LabS start to decrease again, along with sea ice expansion. One possible reason for this reversal could be the strong freshwater influx from the Arctic Ocean through the Denmark Strait, which peaks in phase C and reaches to the LabS. The freshwater flow is concentrated in the upper ocean, implying that it can disrupt convection by freshening the upper water column. Another explanation could involve surface fluxes of heat and freshwater. In phase D, sea ice rapidly expands to cover the entire LabS and convection shuts down, causing the AMOC decline (Figs. \ref{fig:phases}d-g).

Fully deciphering the oscillation mechanism goes beyond the scope of this study. Nonetheless, we can identify multiple competing processes that could produce cyclic behaviour: a competition between sea ice and convection in the LabS, a competition between salt advection by the AMOC and freshwater advection from the Arctic Ocean, as well as a competition of deep water formation sites between the LabS and NorS. The latter could explain the anti-phase pattern observed in precipitation between Greenland and the United Kingdom (Fig. \ref{fig:oscil}g) as well as in surface air temperatures between the Greenland-Iceland-Norwegian (GIN) Seas and the rest of the high northern latitudes (Fig. \ref{fig:phases}c). During phase A, at the AMOC minimum, air temperatures are warmer over the GIN Seas than during phase C, in contrast with the surrounding areas, especially the LabS. We hypothesise that this is because the NorS is the only deep convection zone in the North Atlantic during phase A, directing the meridional heat transport to that region and reducing cold inflows from the Arctic Ocean.

Exploring the dynamics on the edge state thus reveals distinct modes of climate variability that are absent in the ON and OFF attractors -- but become highly relevant near criticality, as we show below. The edge tracking method allowed to capture the centennial climate oscillations even though they are asymptotically unstable. 

\section{Boundary crisis: From edge state to ghost state \label{sec:crisis}}
So far, we have investigated the global stability of the AMOC in PlaSim-LSG at constant external forcing $\bm \lambda$, with the CO$_2$ concentration set to 360\,ppm. However, the radiative forcing of the Earth is currently undergoing rapid change as CO$_2$ concentrations are increasing at a rate of around 0.56\% per year. Consequently, the stability landscape of the Earth system is continuously evolving as a function of $\bm\Lambda(t)$ (see Eq. \eqref{eq:ode}). In this nonautonomous context, attractors and edge states must be viewed in a pullback or snapshot sense as they are moving in state space subject to the change of the control $\bm \Lambda$ \cite{caraballo_applied_2016,wieczorek_rate-induced_2023}.

From bifurcation theory, it is well known that there may be critical forcing levels $\bm \lambda_c$ at which the global stability landscape changes qualitatively. For example, new attractors may emerge, existing ones may disappear, or attractors may switch between periodic and non-periodic behaviour \cite{grebogi_crises_1983}. An important case are \textit{boundary crises} where an attractor is annihilated by colliding with an edge state embedded in a basin boundary \cite{hong_chaotic_2004,axelsen_finite-time_2024}. It has been proposed that their union after the boundary crisis forms a \textit{ghost state} -- a state reminiscent of the dynamics on the former attractor and edge state that has a long mean lifetime yet is asymptotically unstable \cite{feudel_multistability_2018,mehling_limits_2024,koch_ghost_2024}. Any trajectory initialised on the ghost state will eventually diverge from it and approach an attractor, possibly after an ultralong transient. A boundary crisis involving chaotic invariant sets may be viewed as the analogue of a saddle-node bifurcation in non-chaotic systems.

\subsection{AMOC stability landscape as a function of CO$_2$ level}
\begin{figure}
    \centering
    \includegraphics[width=0.9\textwidth]{"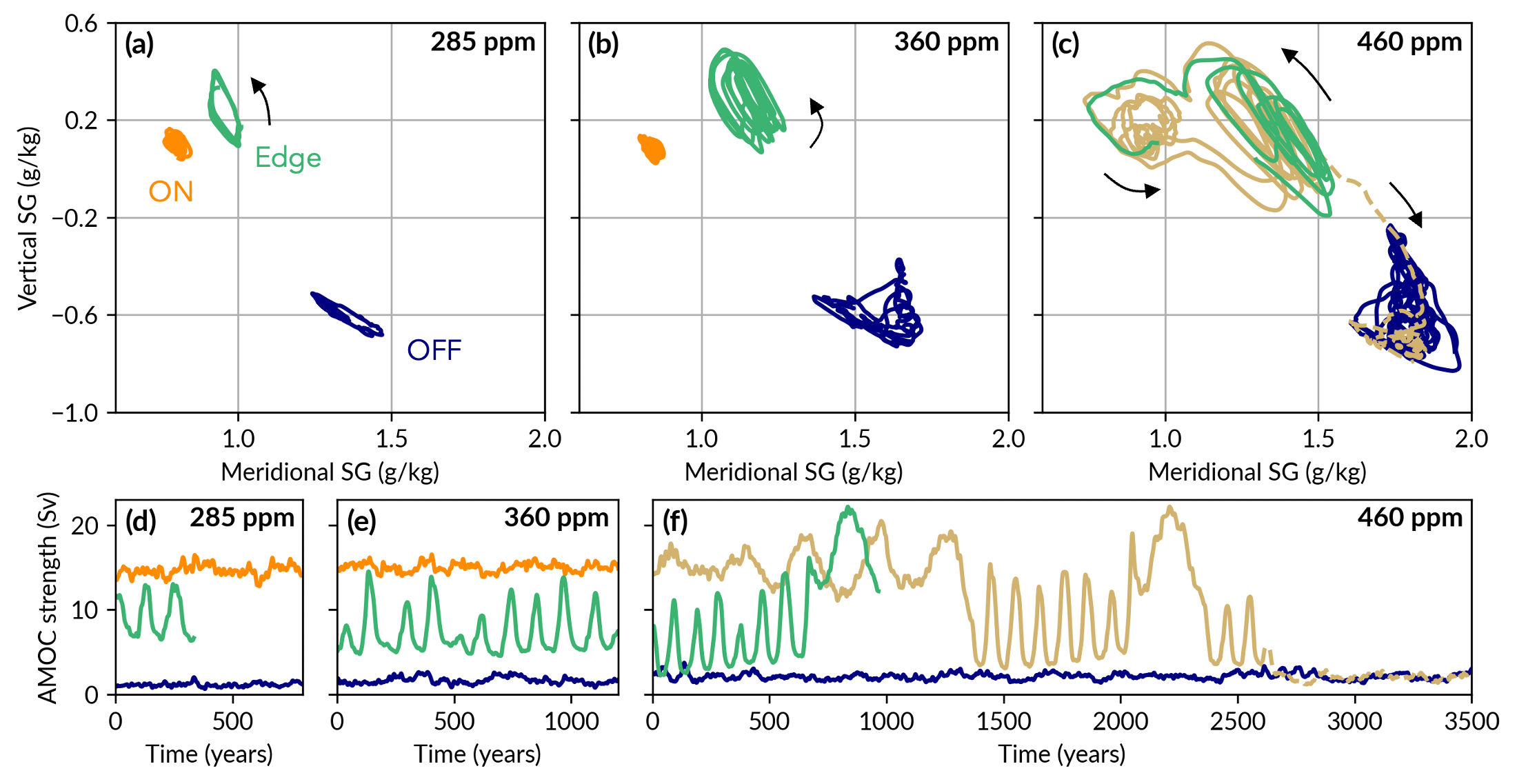"}
    \caption{\label{fig:a2c1states} \textbf{AMOC stability landscape as function of CO$_2$ concentration} for 285\,ppm (a,d),  (b,e) and 460\,ppm (c,f). Reduced state space projections (a-c) onto the meridional and vertical SG show the ON (orange), OFF (blue) and edge trajectory (green). For 460\,ppm, the transient trajectory initialised from the former ON state (beige) and the edge trajectory (green) trace the ghost state. Arrows indicate the time direction. (d-f) AMOC timeseries at 46-66$^\circ$N corresponding to (a-c), respectively.}
\end{figure}

To explore how the stability landscape of the AMOC changes as a function of CO$_2$, we now consider the frozen system (i.e., fixed external forcing) at two additional CO$_2$ levels: 285\,ppm (preindustrial conditions) and 460\,ppm (Fig. \ref{fig:a2c1states}). Analogously to our investigation at 360\,ppm, we run long simulations (4000 years), initialised from the ON and OFF state obtained at 360\,ppm, respectively. Additionally, we run the edge tracking algorithm (sec. \ref{sec:edgetracking}).

In the case of 285\,ppm, the stability landscape qualitatively resembles the situation at : the ON and OFF states, largely unchanged in AMOC strength, are clearly separated from an oscillatory edge state (Fig. \ref{fig:a2c1states}a,d). The edge state oscillations have a similar period but a slightly smaller amplitude compared to . Interestingly, we observed that edge tracking (initialised from the ON and OFF states at 285\,ppm) is considerably more time-consuming at lower CO$_2$, since trajectories tend to diverge more quickly from the basin boundary (40 iterations yielded about 400 years of edge trajectory instead of 1400 years obtained for ). This suggests that the edge state is more repelling at 285\,ppm.

At a higher CO$_2$ level of 460\,ppm, the OFF state persists for the 4000 years of simulation and resembles the OFF state at lower CO$_2$ in terms of AMOC strength (Fig. \ref{fig:a2c1states}f). By contrast, the simulation initialised from the former ON state eventually collapses to the OFF state after a 2700-year long transient. During the first 1300 years, this trajectory (beige line in Fig. \ref{fig:a2c1states}f) maintains a relatively strong AMOC with multi-centennial oscillations reminiscent of those in Ref. \cite{mehling_high-latitude_2022}, growing up to 10\,Sv in amplitude. Then, the AMOC abruptly declines to less than 5\,Sv and enters a period of large oscillations that resemble those of the edge state in period and amplitude. After six cycles, the AMOC suddenly recovers and overshoots to 22\,Sv, thereafter steeply declines again, and eventually collapses to the OFF state where the trajectory remains for the final 1200 years of simulation.

Dynamically, two possible situations could explain this behaviour. The basin boundary at 460\,ppm could have moved in the state space such that initial conditions on the ON state at  now lie in the basin of attraction of the OFF state. Alternatively, the ON state could have disappeared entirely at 460\,ppm, implying a monostable regime with the OFF state being the only asymptotically stable attractor. In the following, we argue for the latter possibility.

\subsection{Collision of ON and edge states}
Despite the fact that the ON state has lost its stability at 460\,ppm, edge tracking between the ON and OFF states is still possible for a while. This is because the former ON state is transiently stable for a few hundred years. We can thus find pairs of initial conditions that converge to a weak and, temporarily, a strong AMOC state, respectively. Running the edge tracking algorithm at 460\,ppm (initialised from the ON and OFF states at ) produces several large AMOC oscillations that resemble the edge state dynamics at 360\,ppm, though the AMOC minimum is initially lower and the period of around 100 years is slightly shorter than at 360\,ppm.

After about 750 years of edge tracking, the edge trajectory interrupts its oscillatory behaviour and follows a course that is characteristic for relaxation paths from the edge state to the former ON state. Seemingly, the edge tracking algorithm loses track of the edge state and instead approaches the ON state. However, we have seen that this \enquote{ON state} is not an attractor anymore. Rather, the former ON and edge states are now an intertwined chaotic object -- a ghost state \cite{feudel_multistability_2018,koch_ghost_2024}.

To see this, let us project the dynamics onto the reduced state space spanned by the meridional and vertical SG (see section \ref{sec:reduced-phase-space}). As CO$_2$ increases, all states shift slightly towards larger meridional SG values (Fig. \ref{fig:a2c1states}). The OFF and edge state display higher variability, taking up an increasing volume in the reduced state space. Strikingly, at 460\,ppm, the former ON state and edge state now extend so much that they \enquote{touch} and are not separated anymore. Ocillations of the edge trajectory extend further to higher meridional and lower vertical SG values compared to the edge state at . Eventually, the edge trajectory transitions to the region of the former ON state, and the trajectory initialised from the former ON state circles around the ON state region and then transitions to the edge state region, where it undergoes the same oscillations as the edge trajectory before moving to the OFF state.

Based on the state space view taken here, we propose that the ghost state embodies the union of two interconnected state space regions with rotational dynamics: multi-centennial oscillations near the former ON state and centennial oscillations near the former edge state. Since both regions are not separated in state space, trajectories can chaotically switch back and forth between both oscillatory modes until they necessarily escape the ghost state and converge to the OFF state. 

\subsection{Chaotic transients \label{sec:transients}}
\begin{figure}
    \centering
    \includegraphics[width=0.85\textwidth]{"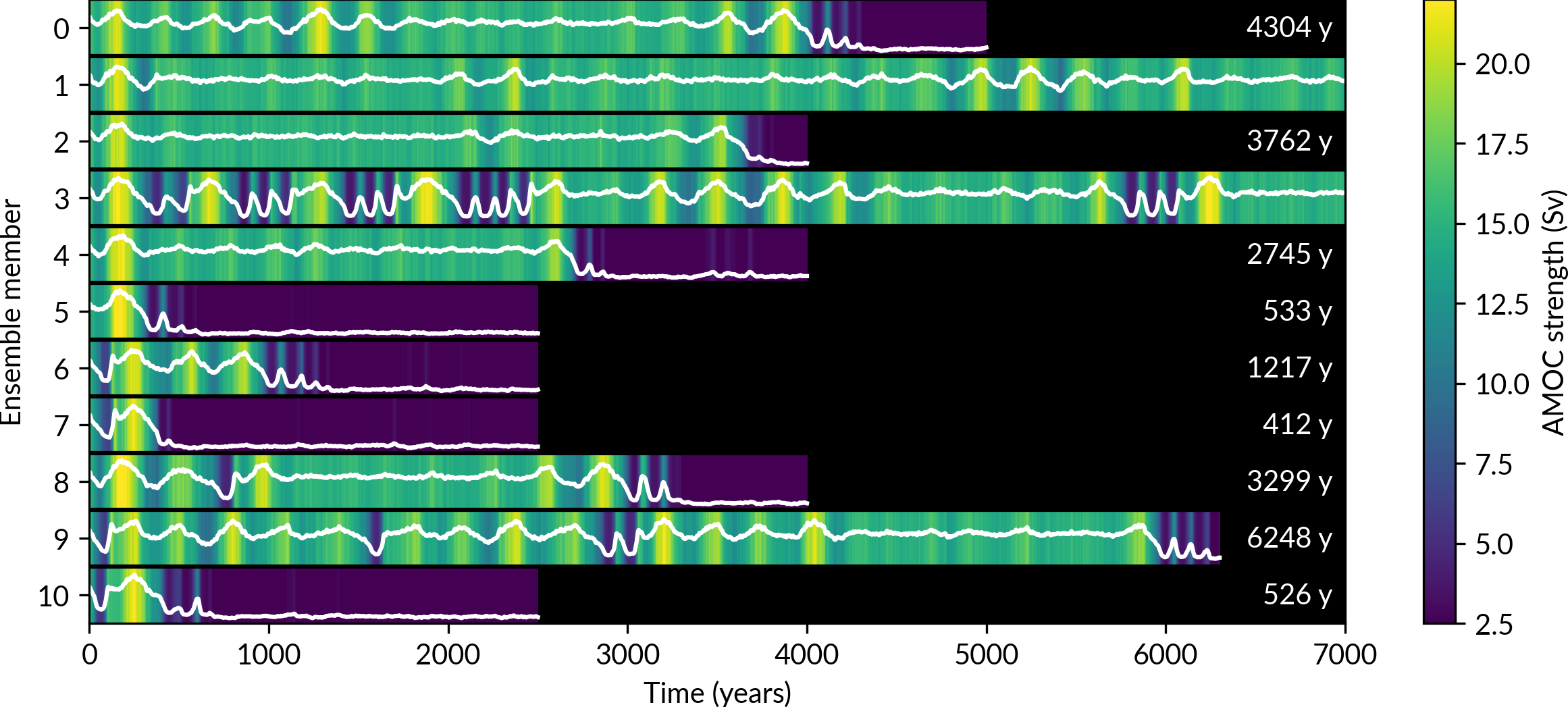"}
    \caption{\label{fig:transients} \textbf{Chaotic transients} in an 11-member ensemble of simulations under constant forcing (460\,ppm CO$_2$), initialised near the ghost state. Each row represents one trajectory coloured by the AMOC strength at 46-66$^\circ$N, stating the duration of the transient (number of years until the AMOC strength first drops below 3\,Sv).}
\end{figure}

To further explore the transient dynamics in the monostable regime, we exploit simulations produced as part of the edge tracking procedure at 460\,ppm. Specifically, we consider the ensemble of 11 simulations used for the last iteration of edge tracking before the edge trajectory jumps to the ON state region (iteration 13). These simulations are run from nearby initial conditions interpolated between one that collapsed (member 0) and another that maintained a strong AMOC (member 10) within 500 years during the previous edge tracking iteration.

The ensemble reveals a rich transient behavior (Fig. \ref{fig:transients}). Initially, all trajectories undergo a spike in AMOC strength corresponding to an excursion to the former ON state region. Thereafter, the AMOC evolution varies greatly between ensemble members. While some members collapse after around 400-500 years, others take over 6000 years before collapsing. In fact, two members do not collapse within 7000 years of simulation. Nonetheless, we expect them to eventually collapse if the simulation would be extended. As seen from members 2 and 9, for example, the collapse can happen relatively abruptly without apparent pre-warning. During the transients, ensemble trajectories exhibit the different patterns of variability associated with the ghost state: slower, less regular oscillations of a stronger AMOC associated with the former ON state, and episodes of more rapid edge state-like oscillations.

This demonstrates that the transient dynamics near the ghost state are essentially unpredictable and can last for thousands of years. The long lifetime of the ghost state suggests that at 460\,ppm our model is close to the boundary crisis where the ON and edge state merged, which occurs somewhere between 360 and 460\,ppm. Further away from the critical CO$_2$ value, the ghost state is expected to have a shorter lifetime (see Ref. \cite{mehling_limits_2024}). Indeed, attempting to perform edge tracking at 500 and 540\,ppm proved unsuccessful because the model quickly diverged from the ON state, as a result of the enhanced instability of the system.

\section{Role of the edge state under nonautonomous climate forcing \label{sec:ssp-forcing}}
Our study has been focusing on snapshots of the stability landscape of PlaSim-LSG at fixed external forcing: we investigated the model as an autonomous dynamical system at different CO$_2$ concentrations.
What can our results tell us about the transition behaviour of the AMOC in a nonautonomous context?

\subsection{State space trajectories under future SSP scenarios}

Let us return to the CO$_2$ forcing experiments introduced at the beginning of this paper (Fig. \ref{fig:ssp-triad}), where we forced PlaSim-LSG with the CO$_2$ projections of low, intermediate, and high emission SSP scenarios. Recall that the AMOC persists under the low emissions scenario, transitions to the OFF state at high emissions, and exhibits a splitting of the simulation ensemble at intermediate emissions.

\begin{figure*}
    \centering
    \includegraphics[width=\textwidth]{"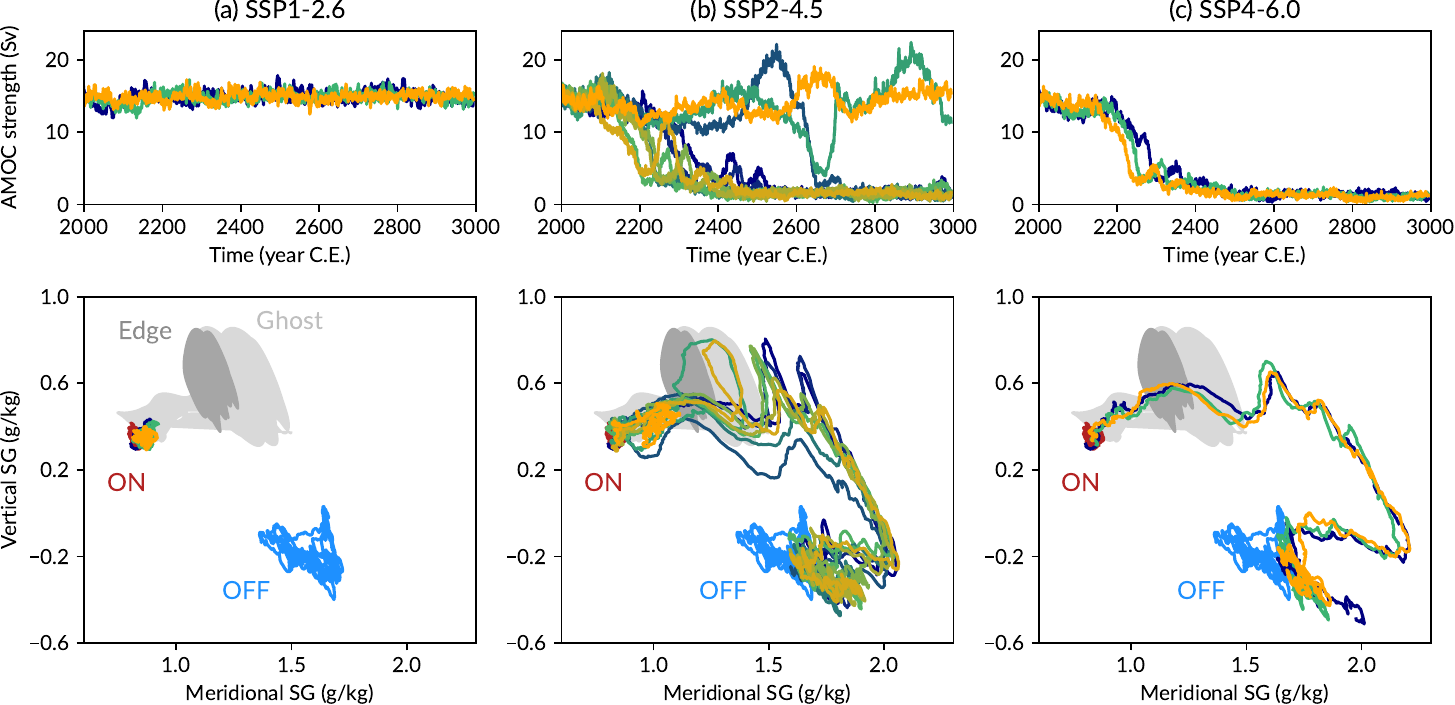"}
    \caption{\label{fig:ssp-proj} \textbf{Trajectories of PlaSim-LSG under future CO$_2$ emission scenarios} for ensemble members run under the (a) SSP1-2.6, (b) SSP2-4.5 and (c) SSP4-6.0 scenario. Top panels show the AMOC timeseries, bottom panels show their projection onto the reduced state space spanned by the meridional and vertical SG. Dark and light gray shaded areas indicate the region of the edge state at  and ghost state at 460\,ppm, respectively. The ON (red) and OFF (blue) states at  are shown for reference.}
\end{figure*}

We can now inspect these simulations in the reduced state space projection to see how their trajectories in state space relate to the model's stability landscape, particularly the edge state and ghost state (Fig. \ref{fig:ssp-proj}). For SSP1-2.6, the ensemble members remain in the region of the ON state. In the SSP4-6.0 scenario, the trajectories pass straight through the ghost state region, as if the oscillatory regime of the ghost state would be \enquote{invisible} to them. For SSP2-4.5, over the 1000-year simulation period, one ensemble member remains to the left of the edge state region, maintaining a strong AMOC; some trajectories travel through the lower part of the edge state region (where the AMOC is weakest, see Fig. \ref{fig:phases}a) and collapse to the OFF state; yet other ensemble members perform one or more cycles of an oscillatory motion before converging to the OFF state. These oscillations occur in the region of the edge and ghost states or to the right of it. Since the states tend to move to higher meridional SG values with increasing CO$_2$ (Fig. \ref{fig:a2c1states}), it is likely that the ghost state likewise expands to higher meridional SG values above 460\,ppm. 

These results indicate that the edge state and, beyond the boundary crisis, the corresponding ghost state play a key role in the ensemble splitting with respect to the AMOC strength, observed under the intermediate CO$_2$ forcing. Under the low emissions scenario, the trajectories do not travel to the edge state region, while under the high emissions scenario the forcing rate is so high that the dynamical structure of the frozen system is masked. Indeed, the SSP1-2.6 scenario remains below 460\,ppm (besides a short overshoot, see Fig. \ref{fig:ssp-triad}a), such that the ON state continues to exist. By contrast, the SSP2-4.5 and SSP4-6.0 scenarios stay above 460\,ppm after the year 2050, such that we assume the model is in the monostable regime from then onwards. 

\subsection{Stochastic bifurcation in the GISS model \label{sec:giss}}
\begin{figure*}
    \centering
    \includegraphics[width=0.8\textwidth]{"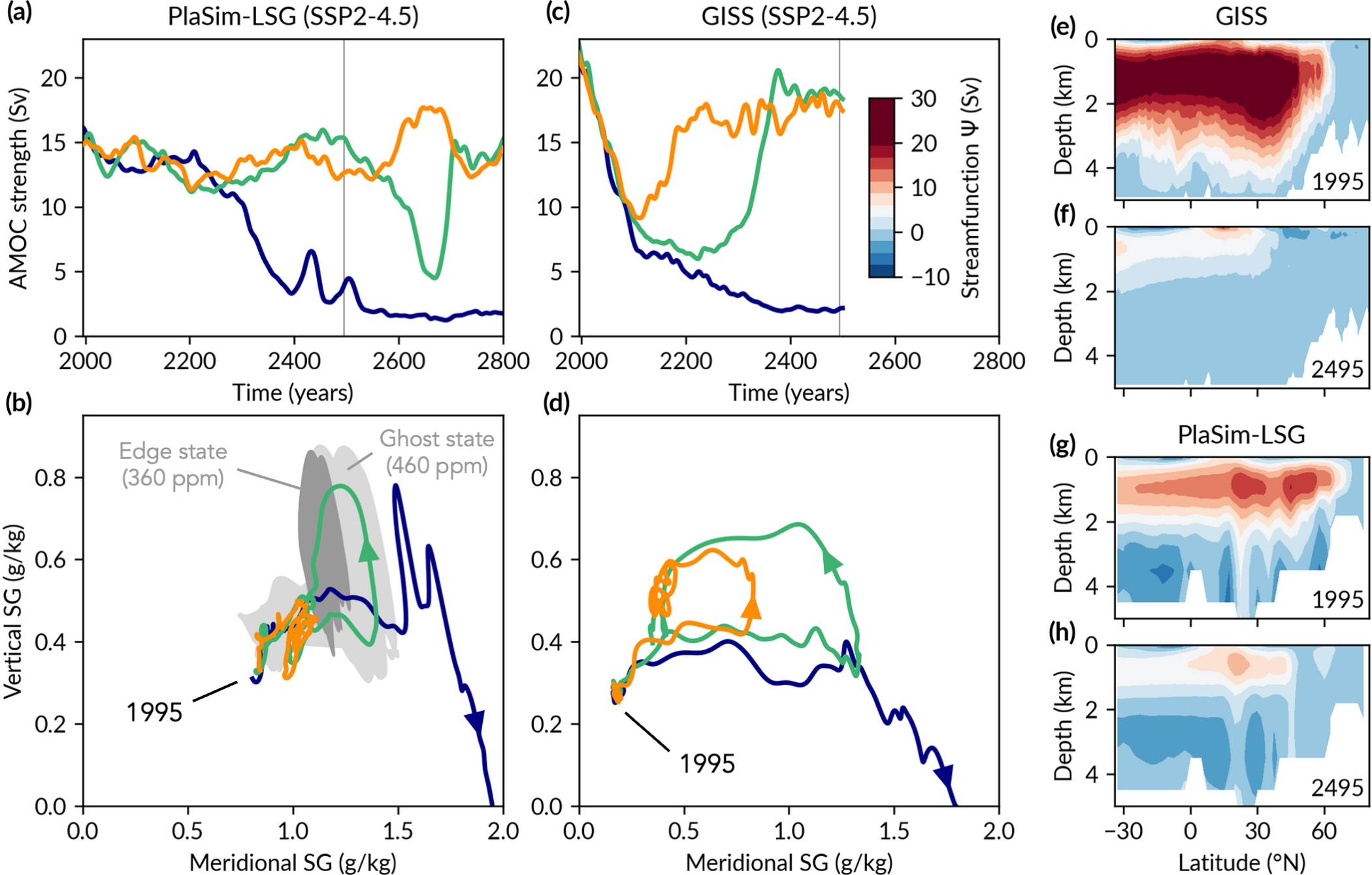"}
    \caption{\label{fig:giss-plasim} \textbf{Comparison between GISS and PlaSim-LSG simulations} under the SSP2-4.5 scenario. (a)-(b) AMOC strength at 46-66$^\circ$N for the three selected ensemble members in each model, couloured by similarity. (c)-(d) State space projection of the trajectories in (a) and (b), respectively, with arrows indicating the time direction. Dark and light gray shaded areas indicate the region of the edge state (at 360\,ppm) and ghost state (at 460\,ppm) found in PlaSim-LSG, respectively. (e)-(h) 5-year averages of the Atlantic meridional streamfunctions along the dark blue trajectories of each model, starting in the years 1995 and 2495 as labeled.}
\end{figure*}

The diverging AMOC behaviour of ensemble members observed in PlaSim-LSG under SSP2-4.5 is intriguingly reminiscent of the so-called \enquote{stochastic bifurcation} found under the same forcing scenario in the CMIP6 model GISS-E2-1-G \cite{romanou_stochastic_2023}. Can the global stability perspective presented here help explain the dynamics in that more comprehensive model?

We select three ensemble members of the GISS model simulations under SSP2-4.5 (members r1i1p1f2, r7i1p1f2, and r10i1p1f2). The simulations extend until the year 2500 and show divergent AMOC behaviour (Fig. \ref{fig:giss-plasim}c). Following an initial AMOC weakening in all members, the first member starts to recover around the year 2100, the second recovers after 2200, whereas the third remains in a weak AMOC state until 2500 (but eventually recovers, see Ref. \cite{romanou_stochastic_2023}). Similar to PlaSim-LSG, the weak AMOC state is characterised by a collapse of the overturning cell north of 45$^\circ$N while a weak overturning circulation is maintained south of 45$^\circ$N (Fig. \ref{fig:giss-plasim}h). In the year 1995 CE, when CO$_2$ levels are at 360\,ppm and the AMOC is in the ON state, the AMOC is around 30\% stronger in GISS than in PlaSim-LSG, though the meridional streamfunctions have a qualitatively similar shape (Fig. \ref{fig:giss-plasim}e, g)).

To relate these simulations to our results, we likewise select three ensemble members from the SSP2-4.5 simulations with PlaSim-LSG, based on their qualitative similarity with each of the GISS ensemble member (Fig. \ref{fig:giss-plasim}a). The first member maintains a strong AMOC, the second undergoes a weakening to about 5\,Sv followed by a recovery, and the third collapses to the OFF state. Note that the evolution of the trajectories is delayed in PlaSim-LSG compared to the GISS model, and the initial AMOC weakening is less pronounced.

We now compare the reduced state space trajectories of these simulations between the two models. Using the same definitions as for PlaSim-LSG, we compute the meridional and vertical SG in GISS based on the Atlantic zonally averaged salinity field. Due to the complexity of the GISS model, its AMOC stability landscape with respect to CO$_2$ and the properties of potential edge states or ghost states are not known. However, we can study how the trajectories relate to the edge state and ghost state found in PlaSim-LSG.

Remarkably, the reduced state space dynamics are qualitatively similar between the two models (Fig. \ref{fig:giss-plasim}b, d). The GISS trajectories start off from a significantly lower meridional SG than the PlaSim-LSG trajectories, in line with the fact that the AMOC in GISS is significantly stronger at that time. The vertical SG values are in good agreement between the models. As CO$_2$ forcing increases, all trajectories initially move towards larger meridional and slightly larger vertical SG values. The AMOC recovery in GISS is characterised by a counter-clockwise rotation, where the loop performs a larger excursion for the trajectory that recovers from lower AMOC values. Interestingly, the reversal of the GISS trajectory with a late recovery occurs directly in the state space region where the edge state is located in PlaSim-LSG, following a path that resembles that of the recovering PlaSim-LSG trajectory. The collapsing trajectory in GISS skims the bottom end of the ghost state region before traveling to high meridional and low vertical SG values. This path is qualitatively similar to the collapsing PlaSim-LSG trajectory. Although the collapsing GISS simulation does not display any AMOC oscillations seen in the collapsing PlaSim-LSG simulation, there are still upward spikes in the state space trajectory that might hint at similar, yet dampened dynamics.

To summarise, we find that the splitting of the GISS ensemble occurs in the same region of the projected state space in which the edge state is located in PlaSim-LSG. This supports the proposition that the \enquote{stochastic bifurcation} could indeed be a signature of a chaotic AMOC edge state near a boundary crisis.

\section{Discussion and conclusion}
This paper presents a global view of the stability landscape of the AMOC in a coupled climate model. While mapping out the full quasipotential landscape seems out of reach for a 10$^5$-dimensional system, we present a proof of concept that analysing edge states gives key insights into the global stability properties, transient dynamics and instability mechanisms of high-dimensional climate models.

Traditionally, studies of climate tipping points often focus on the local dynamics near stable equilibria. Particularly, statistical EWS based on critical slowing-down measure changes in the local stability of an attractor under a quasi-adiabatic parameter drift: as the system approaches a bifurcation, the quasipotential flattens around the attractor, implying a reduction in restoring forces. Another way to look at this is that the barrier imposed by the edge state diminishes towards the bifurcation, and that the quasipotential flattens around the edge state (with an opposite sign of the curvature). This fits to our observation that edge tracking was more expensive at lower CO$_2$ concentrations, further from the boundary crisis. Closer to the crisis, the edge state becomes \enquote{stickier} \cite{lai_algebraic_1992,lai_transient_2011} in the sense that trajectories tend to spend longer times in its vicinity, suggesting an alternative, non-local angle on critical slowing-down\footnote{Ying-Cheng Lai, personal communication.}. It seems clear that the current rate of anthropogenic greenhouse gas emissions is forcing the climate system out of a steady-state. Our results indicate that the edge state dynamics can become relevant under plausible future emissions scenarios: the boundary crisis in PlaSim-LSG occurs at CO$_2$ levels that could be reached within two decades. Nonetheless, whether this boundary crisis is a feature of the real climate system remains unknown.

This study was conducted with a climate model of intermediate complexity that inevitably relies on simplifications and neglects numerous processes of potential relevance. Therefore, the results of our investigation may be highly model-dependent and not representative of reality. Even though the AMOC edge state found in PlaSim-LSG is a physically sensible steady-state, its nonlinear dynamics might be exaggerated effects of the highly simplified parameterisations of, e.g., sea ice and oceanic convection. 
On the other hand, we believe that PlaSim-LSG is to date the most complex climate model in which an edge state has been explicitly computed. Our results thus add a significant step towards realism to recent studies investigating edge states of the AMOC in a conceptual climate model \cite{mehling_limits_2024} and a global ocean-only circulation model \cite{lohmann_melancholia_2024}. Furthermore, the similarity in the dynamics between PlaSim-LSG and the more complex GISS model suggests that the global stability view established here could provide key insights into the behaviour of state-of-the-art earth system models. As is increasingly clear, AMOC metastability and tipping behaviour is not restricted to simple climate models but occurs across the model hierarchy \cite{romanou_stochastic_2023, van_westen_asymmetry_2023, cini_simulating_2024, willeit_generalized_2024}. The state space and parameter space of large models are just more challenging to explore.

The key limitation for applying the edge tracking algorithm in even higher-dimensional systems is the computational cost of running long simulations. In PlaSim-LSG, producing one year of edge trajectory required on average 50 (90) years of simuation at  (460\,ppm). Thus, the 1400 year-long edge trajectory in the  case consumed around 70\,000 simulation years or 3000 CPU hours. Of course, this number depends strongly on the system and could be optimised by tuning the settings of the edge tracking procedure.

The edge tracking algorithm converges to an edge state despite the complex geometry of the basin boundary, which is typically fractal \cite{bodai_rough_2020, lucarini_edge_2017, mehling_limits_2024}. In our study, we observed that the basin boundary is folded along transects of interpolated initial conditions, hinting at fractality. Due to the long lifetime of the ghost state, edge tracking appears to work for multiple iterations even beyond the boundary crisis (in the monostable regime). This permits probing ghost states while also demonstrating the difficulty of precisely determining critical forcing thresholds \cite{mehling_limits_2024}.
From a modelling perspective, it is not obvious that the interpolation between initial conditions in all dynamical variables yields new initial conditions that generate numerically stable and physical trajectories. We argue that convexity of the equations governing the climate dynamics ensures that trajectories quickly relax to a physical state. 

Using an ocean-only model, Ref. \cite{lohmann_melancholia_2024} found that the AMOC edge state features a less \enquote{spicy} (i.e., colder and fresher) deep North Atlantic than the attractors, as well as a higher dynamic enthalpy. Our results corroborate this in a coupled climate model while also revealing a much richer dynamics due to the ice-ocean-atmosphere coupling, in which the upper ocean plays a more active role. Whereas Ref. \cite{lohmann_melancholia_2024} concluded that the most relevant regions for anticipating AMOC transitions are located in the deep sea, our results suggest that many excursive observables are found also in the surface ocean. This could potentially be exploited for improved early warning systems of AMOC changes \cite{lohmann_role_2024,lenton_remotely_2024}.

AMOC oscillations have received wide interest due to their occurrence in various climate models and potential for explaining past abrupt climate change. In the context of Dansgaard-Oeschger events \cite{dansgaard_evidence_1993}, previous work has determined \enquote{sweet spots} for oscillations in parameter space \cite{malmierca-vallet_impact_2024}. We demonstrate a sweet spot in state space: while the ON and OFF states do not exhibit oscillations at , oscillations occur near the edge state. The drivers of these unstable oscillations involve similar processes previously identified in stable oscillation mechanisms in other models \cite{li_coupled_2019,mehling_centennial-scale_2024,rome_simulated_2025}. 
Dynamically, the presence of unstable oscillations near the crisis might hint at the existence of a subcritical Hopf bifurcation with respect to CO$_2$. Oscillations in the AMOC strength and other popular observables such as the freshwater transport into the Atlantic further suggest that such quantities may be poor indicators of AMOC stability in out-of-equilibrium conditions.


The limited predictability of the AMOC near an instability has already been suggested by Ref. \cite{knutti_limited_2002}. Refs. \cite{lohmann_predictability_2024,romanou_stochastic_2023} have recently reiterated this idea by demonstrating an ensemble splitting caused by internal variability under identical time-dependent forcing. 
Our findings allow to understand this behaviour in terms of an edge state and, beyond the boundary crisis, a ghost state. We can thus directly link the dynamics of earth system models to fundamental concepts of dynamical systems theory that are often only explored in low-dimensional systems.

\paragraph*{Data availability.} Selected simulation data and source code for implementing the edge tracking algorithm in PlaSim-LSG are available at \url{https://doi.org/10.5281/zenodo.17053348}. Further raw model simulation output can be provided by the authors upon request.

\paragraph*{Acknowledgements.}{The authors would like to thank P. Ashwin, K. Bellomo, M. Cini, S. Corti, H. Dijkstra, J. Lohmann, F. Ragone, A. Romanou, T. Tél and S. Wieczorek for useful discussions, and an anonymous reviewer for a fruitful suggestion. All authors acknowledge funding from the European Union's Horizon 2020 research and innovation programme under the Marie Skodowska-Curie Grant Agreement No. 956170 (CriticalEarth). VL acknowledges financial support received from the Horizon Europe projects ClimTip (Grant No. 101137601) Past2Future (Grant No. 101184070), from the ARIA project AdvanTip, and from the ESA project PREDICT. The authors gratefully acknowledge the World Climate Research Programme for coordinating CMIP6, the NASA Goddard Institute  for Space Studies for producing and sharing the output of the GISS model, and the
Earth System Grid Federation (ESGF) for
archiving the data and providing free access.}

\bibliography{main}

\clearpage
\setcounter{equation}{0}
\setcounter{table}{0}
\setcounter{figure}{0}
\setcounter{section}{0}

\renewcommand{\theequation}{S\arabic{equation}}
\renewcommand{\thetable}{S\arabic{table}}
\renewcommand{\thefigure}{S\arabic{figure}}
\renewcommand{\thesection}{S\arabic{section}}

{\Huge{\centering Supplemental Information}}

\section{Model configuration}
We use the version of PlaSim-LSG publicly available at \url{https://github.com/jhardenberg/pLASIM}, with selected namelist parameters specified in Tab. \ref{tab:params}. Complete information on the run settings is provided in the \texttt{DIAG} file included in the data repository (see section \ref{sec:data}).

The vertical diffusivity in the ocean is parameterized according to the profile,
\begin{align}
    A_v (z) = a^* + a_\textbf{range} \arctan \left( \lambda(z-z^*) \right) \,,
\end{align}
where the parameter values for $a^*$ (\texttt{astar}), $a_\text{range}$ (\texttt{arange}), $\lambda$ (\texttt{lambda}) and $z^*$ (\texttt{zstar}) are given in Tab. \ref{tab:params}.

\begin{table}[hb]
\begin{tabular}{llll}
\hline
Parameter name & Description & Value & Unit \\
\hline
\texttt{NFIXORB} & Switch to fix orbital parameters & 1 &  \\
\texttt{ECCEN} & Eccentricity & 1.67$\times 10^{-2}$ &  \\
\texttt{MVELP} & Longitude of perihelion & 102.9 &  \\
\texttt{OBLIQ} & Obliquity & 23.44 &  \\
\texttt{GSOL0} & Solar constant & 1367.0 & W\,m$^{-2}$ \\
\texttt{CO2} & Atm. CO$_2$ concentration & is varied & ppm \\
\texttt{zstar} &  & 2500 & m \\
\texttt{lambda} &  & 4.5$\times 10^{-3}$ & m$^{-1}$ \\
\texttt{astar} &  & 0.8714$\times 10^{-4}$ &  \\
\texttt{arange} &  & 0.2843$\times 10^{-4}$ & \\
\hline
\end{tabular}
\label{tab:params}
\caption{PlaSim-LSG model settings for selected namelist parameters.}
\end{table}

\section{Model diagnostics}
All model output analysed in this study has been converted to annual mean data (from raw data with monthly resolution) before further use. We compute diagnostics in the following way:
\begin{itemize}
    \item \textbf{Density} is computed from salinity, potential temperature and depth using the simplified equation of state (EOS) based on Ref. \cite{vallis_atmospheric_2017}, a nonlinear second-order EOS (see \url{https://www.nemo-ocean.eu/doc/node31.html}).
    \item Since convection is parameterised via a convective adjustment scheme, we must choose a way of estimating the \textbf{convection depth}. For each horizontal grid point, we start at the sea surface and descend until reaching a vertical level for which the annual mean of convective adjustment events is zero. The depth of the previous level (where annual mean convection is nonzero) is taken as the convection depth.
    \item We define the \textbf{sea ice border} as the boundary of the region where the annual mean sea ice thickness is at least \SI{5}{cm}.
\end{itemize}

\section{Edge tracking algorithm}
In PlaSim-LSG, we perform the $k$-th iteration of the edge tracking algorithm in the following way, starting with the initial conditions $\bm x_a^{(0)}$ and $\bm x_b^{(0)}$:
\begin{enumerate}
    \item Compute interpolated initial conditions $\bm x_j^{(k)} = \bm x_a^{(k-1)} + 0.1j\left(\bm x_b^{(k-1)} - \bm x_a^{(k-1)}\right)$ for $j=0,\dots,10$.
    \item Run parallel simulations from $\bm x_j^{(k)}$ until every trajectory can be labelled as approaching either the ON or the OFF state (e.g. around 350 model years at 360\,ppm CO$_2$).
    \item Find new indices $m, n \in \{ 0, \dots, 10\}$ such that the trajectories from $\bm x_m^{(k)}$ and $\bm x_n^{(k)}$ stay close to each other ($<\SI{1}{Sv}$ difference in smoothed AMOC strength) for as long as possible but evolve to different attractors.
    \item Select the time $t_k$ at which the trajectories from $\bm x_m^{(k)}$ and $\bm x_n^{(k)}$ first diverge by $\SI{1}{Sv}$ and use their states at $t_k$ as new initial conditions $\bm x_a^{(k)}$ and $\bm x_b^{(k)}$.
    \item Increase $k$ by 1 and repeat step 1.
\end{enumerate}

\section{Data and software \label{sec:data}}
Selected simulation data and source code for implementing the edge tracking algorithm in PlaSim-LSG are available at \url{https://doi.org/10.5281/zenodo.17053348}. Further raw model simulation output can be provided by the authors upon request.

A general implementation of the edge tracking algorithm in the Julia language is available as part of the software package Attractors.jl \cite{datseris_framework_2023} (\url{https://github.com/JuliaDynamics/Attractors.jl}).

\section{Supplemental figures}

\begin{figure}[h]
    \centering
    \includegraphics[width=0.9\textwidth]{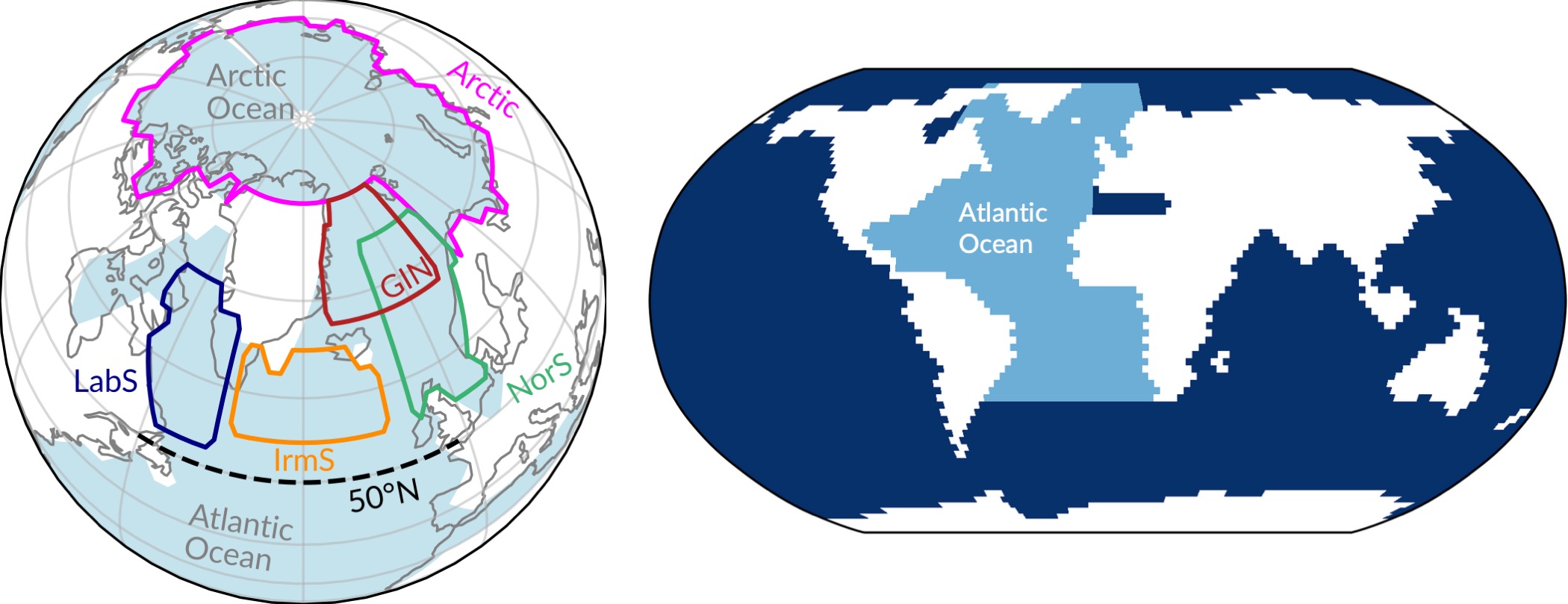}
    \caption{Left: Map showing the model geography in the North Atlantic and the regions defined in this study. Right: World map illustrating the Atlantic basin mask.}
    \label{fig:regions}
\end{figure}

\begin{figure}
    \centering
    \includegraphics[width=0.55\columnwidth]{"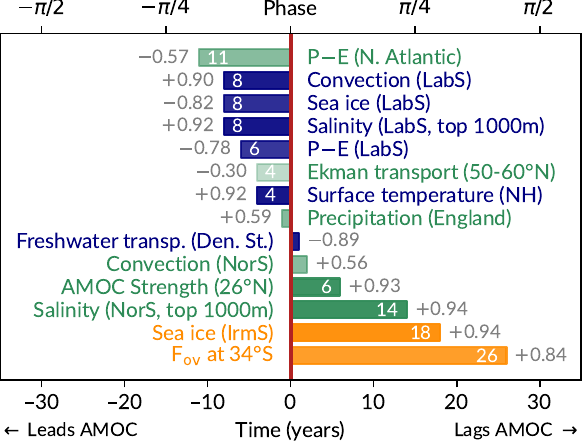"}
    \caption{\label{fig:lags} \textbf{Lag correlations with AMOC strength} for the timeseries shown in Fig. 9 (main text). A negative (positive) lag time means the signal is leading (following) the AMOC at 46-66$^\circ$N. White numbers inside the bars indicate the lag time; gray numbers give the correlation with the AMOC timeseries at that lag (no smoothing, annual resolution).}
\end{figure}

\begin{figure*}[h]
    \centering
    \includegraphics[width=\textwidth]{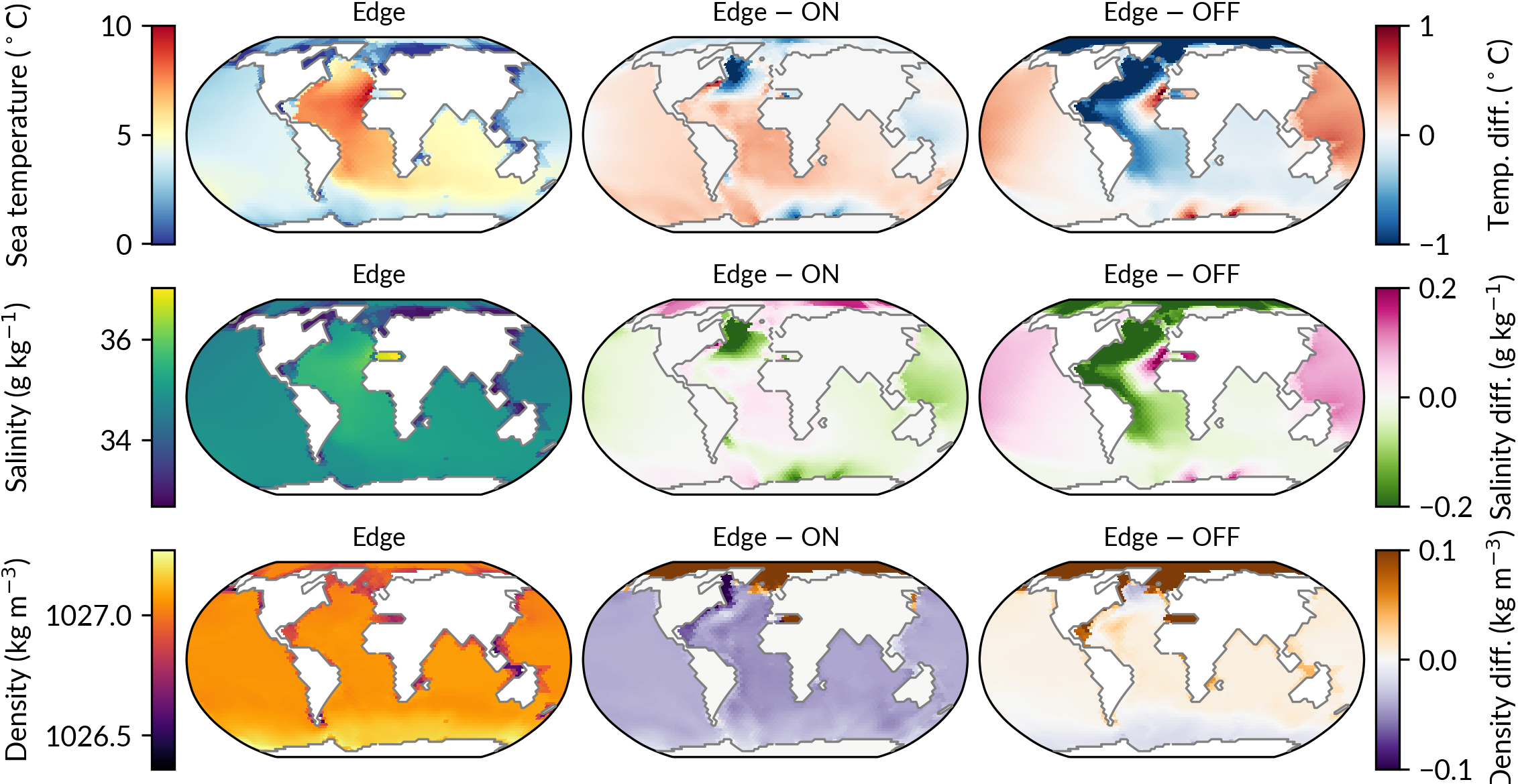}
    \caption{\textbf{Deep sea properties of the edge state} at 360\,ppm CO$_2$. Same as the first three rows of Fig. 8 (main text), but for the deep ocean (averaged over all depths between 1000 and 3000\,m).}
    \label{fig:edge_deep}
\end{figure*}

\begin{figure*}[h]
    \centering
    \includegraphics[width=0.8\textwidth]{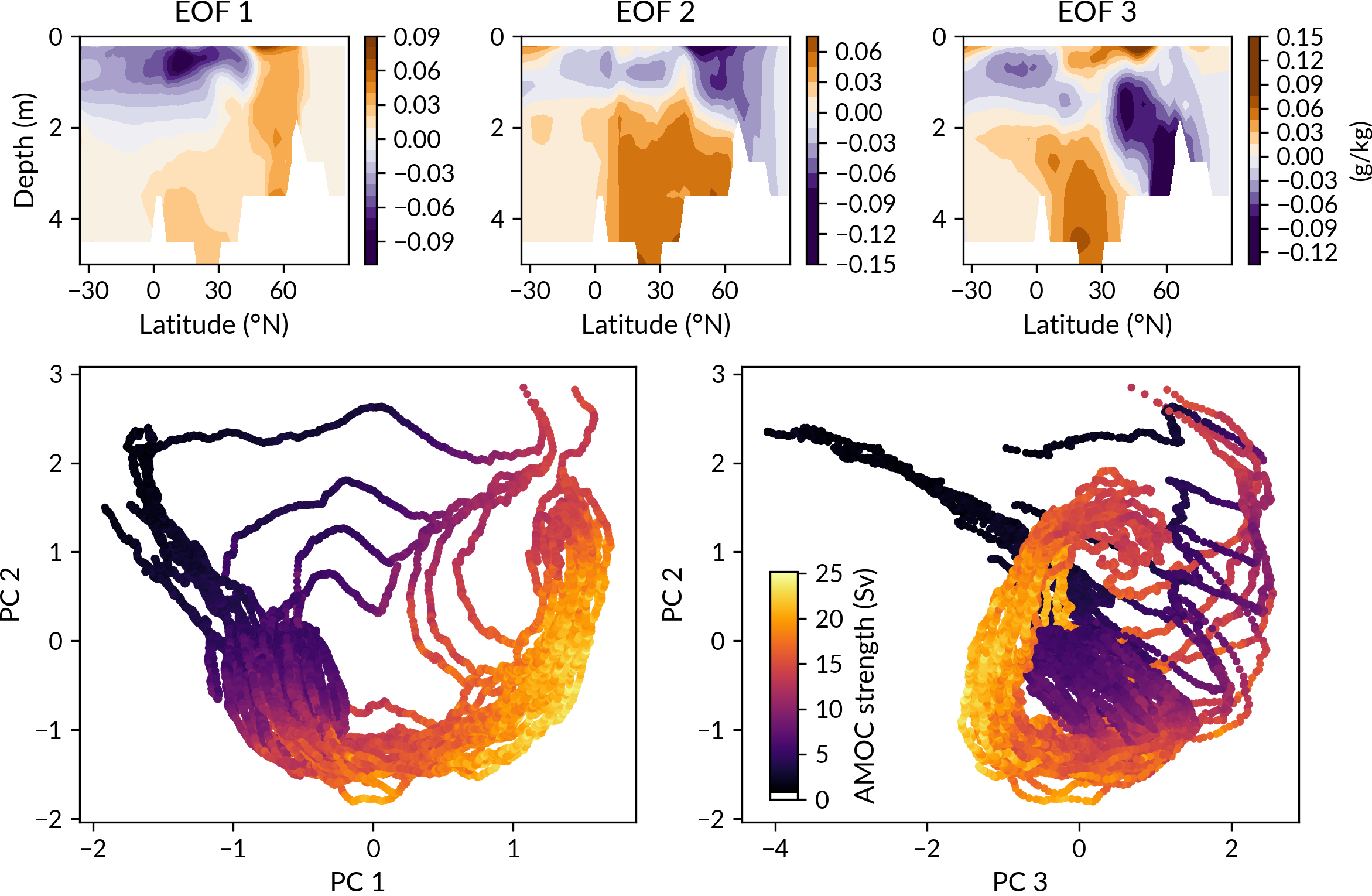}
    \caption{\textbf{EOF analysis} of the edge tracking simulations at 360\,ppm CO$_2$, performed with the \texttt{eofs} package in Python \cite{dawson_eofs_2016}. Top panels show the first three EOFs of the zonally averaged Atlantic salinity field for the simulation data summarized in Fig. \ref{fig:eofs-samples}. Bottom panels show different PCs plotted against each other, colored by AMOC strength along the trajectories.}
    \label{fig:eofs}
\end{figure*}

\begin{figure*}[h]
    \centering
    \includegraphics[width=0.8\textwidth]{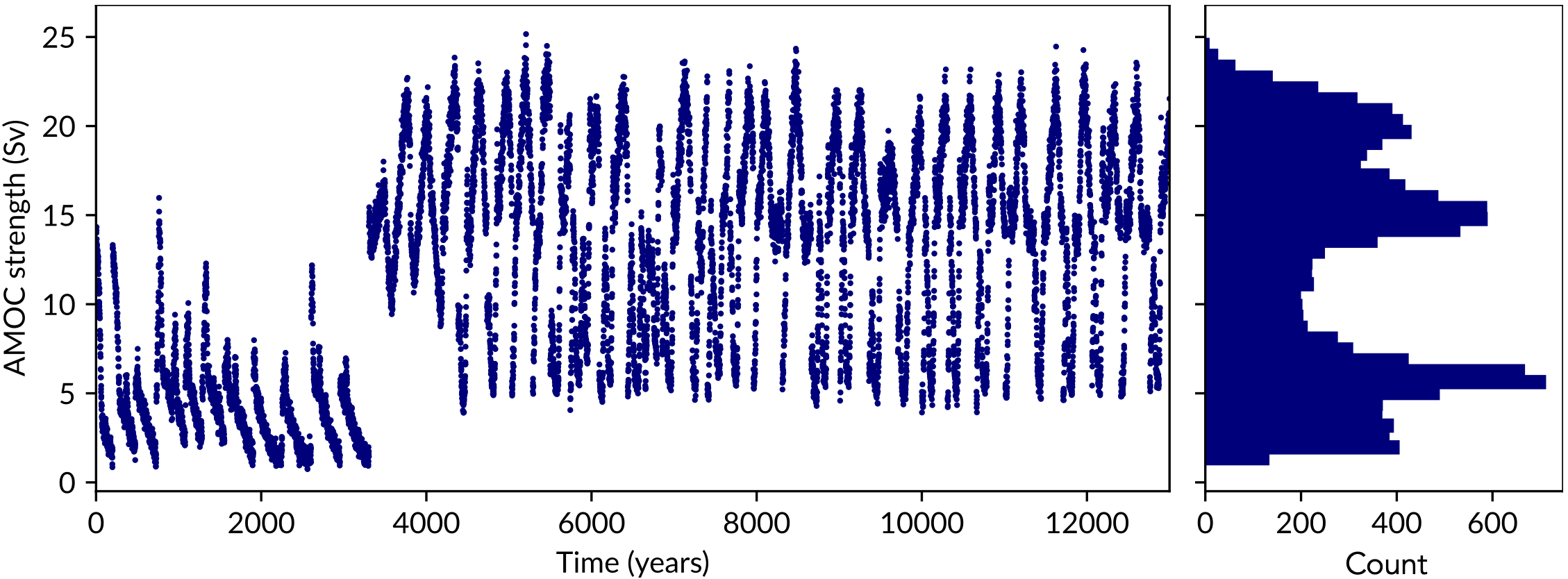}
    \caption{\textbf{Edge tracking simulations used for the EOF analysis} shown in Fig. \ref{fig:eofs}. Left: Concatenated AMOC timeseries for all simulations used. Right: Histogram of AMOC strength across all data points (annual means).}
    \label{fig:eofs-samples}
\end{figure*}

\begin{figure*}[h]
    \centering
    \includegraphics[width=0.55\textwidth]{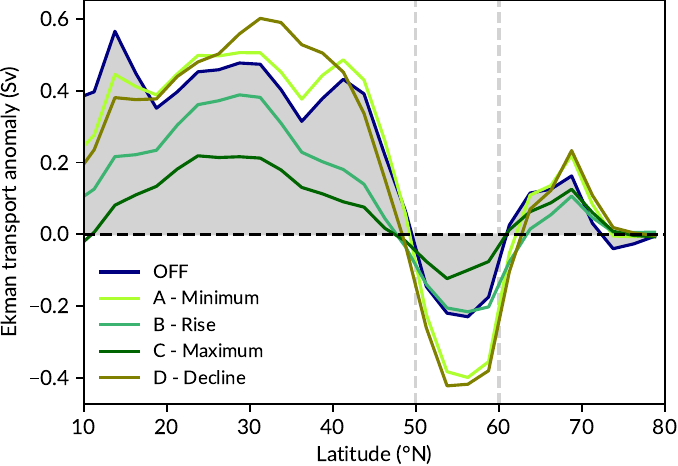}
    \caption{\textbf{Anomaly of the Ekman transport} in the Atlantic basin relative to the ON state, averaged zonally and over time. Positive values indicate a northward transport anomaly. Anomalies are shown for the OFF state (blue, grey shading) and for the different phases A-D on the Edge state (green hues, see main text). Note that between 50-60$^\circ$N, the southward transport anomaly during phases D and A is about twice as strong on the Edge state compared to the ON state.}
    \label{fig:ekman-latitude}
\end{figure*}

\end{document}